\documentclass[a4paper,11pt]{article}
\usepackage{amsfonts}

\usepackage{graphicx}
\usepackage{amsmath}
\usepackage{amssymb}
\usepackage{latexsym}
\usepackage[colorlinks]{hyperref}
\usepackage{color}
\usepackage{float}
\usepackage{cite}
\usepackage{makeidx}
\usepackage{colortbl}
\usepackage{csquotes}
\usepackage[squaren]{SIunits}

\usepackage[textfont={footnotesize,sf},labelfont={color=blue,bf,sf},labelsep=endash]{caption}
\usepackage[position=top,labelfont={color=blue,bf,sf}]{subfig}
\usepackage{bm}

\newcommand{\bra}{\begin{array}}
\newcommand{\era}{\end{array}}
\newcommand{\beq}{\begin{equation}}
\newcommand{\eeq}{\end{equation}}
\newcommand{\bqr}{\begin{eqnarray}}
\newcommand{\eqr}{\end{eqnarray}}

\def\BC{\bb C}
\def\_\BC{\bbi C}

\def\( {\left(}
\def\) {\right)}
\def\no2 {{\textstyle{n\over 2}}}

\def\dag {{\dagger}}

\newcommand{\lga}{\longrightarrow}

\newcommand{\lb}{\label}

\begin{document}

\begin{titlepage}
\setcounter{page}{1}
\renewcommand{\thefootnote}{\fnsymbol{footnote}}

\begin{flushright}
\end{flushright}

\vspace{5mm}

\begin{center}
{\Large \bf {
Electronic Structure of Graphene with two Strains\\
and Double Barrier
}}

\vspace{5mm}

{\bf El Bou\^azzaoui Choubabi}$^{a}$,
{\bf Ahmed Jellal\footnote{\sf 
a.jellal@ucd.ac.ma}}$^{a,b}$, {\bf Abdellatif Kamal}$^{a}$
and {\bf Hocine Bahlouli}$^{b,c}$

\vspace{5mm}

{{$^{a}$\em Laboratory of Theoretical Physics,  
Faculty of Sciences, Choua\"ib Doukkali University},\\
{\em PO Box 20, 24000 El Jadida, Morocco}}

{$^b$\em Saudi Center for Theoretical Physics, Dhahran, Saudi
Arabia}

{$^c$\em Physics Department,  King Fahd University
of Petroleum $\&$ Minerals,\\
Dhahran 31261, Saudi Arabia}

\vspace{3cm}

\begin{abstract}

We study 
the electronic structure of 
%
Dirac fermions scattered by  double barrier potential
in  graphene under strain effect. We 
show that traction and compression strains can be used
to generate 
fermion beam collimation, 1D channels, surface states
and confinement. The corresponding transmission probability
and conductance at zero
temperature are calculated and their numerical 
implementations  taking into account
different configurations of
physical parameters enabled us to analyze some features of 
the system.

\end{abstract}
\end{center}
\vspace{3cm}
\end{titlepage}

\section{Introduction}

Recently,  investigations of the
strain effect on the electronic band structure, optical properties,
electronic transport, spin transport and valley transport in
graphene nanostructures and devices
has attracted much attention from theoretical and experimental
sides \cite{6,7,8,9,10}. These studies were based on
generalized Hamiltonian describing
the quasiparticles in  graphene subject to arbitrary strain effect.
In particular, it turned out that the linear elasticity theory applied to graphene is a reasonable approximation
for relatively small deformations, since    mechanical properties demonstrate
that  graphene can sustain a linear elasticity regime
up to  $20\%$ \cite {3,4,5}.

\begin{figure}[!ht]\centering
  \centering
  \includegraphics[width= 14cm, height=9cm]{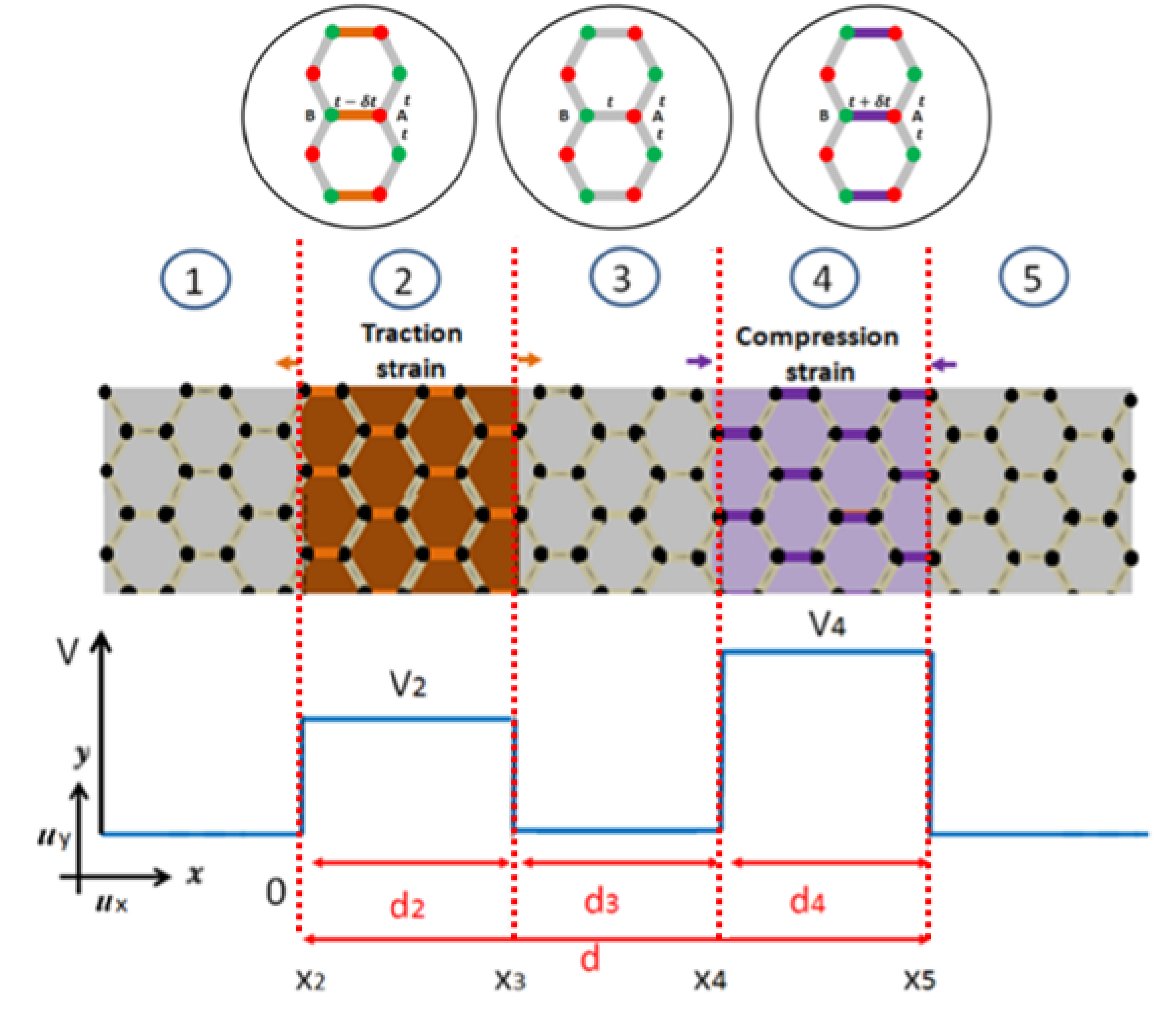}
  \caption{(color online) Inter-carbon links for traction and compression strained graphene
  in the tight-binding model with $\delta t_2=-\delta t_4=\delta t$.}\label{Naj000}
\end{figure}

We intend to study the electronic structure of graphene ribbon deposited on a controllable silicate
dioxide substrate. Such ribbon is of infinite length along the propagating $x$-direction
of  Dirac fermions with 
armchair boundaries (Figure \ref{Naj000}), which
is subject  to two mechanical strains in two separate regions such that strain
results in a contraction in one region and extension in the other. We can achieve this setting by depositing graphene
onto substrates  that allow the control of the strain effect \cite{8}.

\noindent
In the framework
of the tight-binding  model, Figure \ref{Naj000} shows the effect of modifying the horizontal inter-carbon bonds in the ribbon
due to changes in nearest-neighbor hopping amplitude $t$ ($\approx 3\,\text{eV}$).
The space-dependent compression and contraction influence the hopping amplitude through
introducing  modulations
\beq
t(\bm{R}_{i},\bm{n}) = t + \delta t(\bm{R}_{i},\bm{n}), \qquad
t(\bm{R}_{i},\bm{n}) = t - \delta t(\bm{R}_{i},\bm{n})
\eeq
where $\delta t(\bm{R}_{i},\bm{n})$ is a perturbation, $\bm{n}$ is the nearest neighbor vectors
and $\bm{R}_{i} $ the $i$-th site position.
 The ribbon becomes subdivided into five different regions that we index  according to the positive propagation direction as
 $\textbf{\textcircled{1}}$, $\textbf{\textcircled{2}}$, $\textbf{\textcircled{3}}$, $\textbf{\textcircled{4}}$
 and $\textbf{\textcircled{5}}$, successively.
In addition to this ribbon configuration, we apply three different potentials $V_{2}$, $V_{3}$ and $V_{4}$
 in the three regions $\textbf{\textcircled{2}}$, $\textbf{\textcircled{3}}$
and $\textbf{\textcircled{4}}$ of width $ d_2 $, $ d_3$ and $ d_4 $, respectively.
The \emph{input} and \emph{output} regions, $\textbf{\textcircled{1}}$ and $\textbf{\textcircled{5}}$,
are  made of  pristine graphene and
the potential is zero. 
The strained graphene ribbon and the profile potential constitute
a double barrier of width $d=d_2+d_3+d_4$ as indicated in Figure \ref{Naj000}.

The present paper is organized as follows. In section 2, we set the theoretical model
and solve the eigenvalue equation to obtain the solutions of the energy spectrum.
We use the boundary conditions together with the continuity equations to determine
the transmission and reflection probabilities as well as
the conductance in section 3. We numerically analyze 
 the main results and present different discussions in section 4. We conclude our work in the final section.

\section{Theoretical model}

In  strained graphene, near the valleys $K$ and $K'$ of the  first Brillouin zone,
the modification of hopping energies between different sites, through smooth perturbations,
is governed by the following low energy Hamiltonian  \cite{Castro}
\begin{equation}
  H = v_F \int d\textbf{r} \Psi^\dagger
  \begin{bmatrix}
\bm{\sigma} \cdot \left(\bm{p}-\frac{1}{v_F}\bm{\mathcal{A}} \right) & 0 \\
0 & -\bm{\sigma}\cdot \left(\bm{p}+\frac{1}{v_F}\bm{\mathcal{A}}\right)
\end{bmatrix}\Psi
\end{equation}
with   the momentum operator $\bm{p}=(p_x,p_y)$,
the Fermi velocity $v_F=3ta/2\hbar\approx 10^6 m/s$, the Pauli matrices $\bm{\sigma}=(\sigma_x,\sigma_y,\sigma_z)$,
the gauge field $\bm{\mathcal{A}}$, acting on the  electrons dynamics  described by a
Dirac equation, and the spinor
$\Psi=\left[\psi^A_K(\bm{r}),\,\psi^B_K(\bm{r}),\,\psi^B_{K'}(\bm{r}),\,\psi^A_{K'}(\bm{r})\right]^\dagger$.
The gauge field comes from the perturbation of homogeneous amplitude jump $\delta t(\bm{R},\bm{n})$
 and they are related via
\begin{equation}
  \bm{\mathcal{A}}(\bm{r})=\mathcal{A}_x(\bm{r})-i\mathcal{A}_y(\bm{r})
  = \sum_{\bm{n}}\delta t(\bm{r},\bm{n})\: e^{i\bm{K}\cdot \bm{n}}.
\end{equation}
In the double barrier system,  regions $\textbf{\textcircled{2}}$ and
$\textbf{\textcircled{4}}$ have the  same width $\varrho=d_2=d_3=d_4$, the perturbations
$\delta t_j$ of horizontal  hopping  are constant, the index $j$ labels the different regions
and runs from $1$ to $5$. Such perturbations and the associated
$ \bm{\mathcal{A}}_j$$(\bm{r})$ can be written as
\begin{subequations}\label{eq:01}
\begin{align}
  &\delta t_{j}(\bm{R_i},\bm{n}) = \delta t_{j}\, \delta_{\bm{n},0}\, \theta(X_i-x_j)\theta(w-X_i+x_j)
  \label{eq:StepPerturbation-a}
  \\
  &\bm{\mathcal{A}}_j(\bm{r}) = \delta t_j\, \theta(x-x_j) \theta(w-x+x_j) \bm{u}_y
  \label{eq:StepPerturbation-b}
  \,.
\end{align}
\end{subequations}
Note that
 the unit vector $\bm{u}_y$ is collinear to the gauge field $\bm{\mathcal{A}}_j$
 and   is perpendicular to the propagating  $\bm{u}_x$-direction. We can generalize
 the Hamiltonian to describe all regions composing our system with  different couplings. Then adopting the unit system
$v_F =\hbar=1$ and allowing for the presence of an
electrostatic potential $V_j(x,y)$ in the barrier region \cite{8},
the wave equations for the $K$ valley can then be cast into 
\begin{subequations}\label{eq:0001}
\begin{align}
 \bigl[-i(\partial_x-\mathcal{A}_{yj}(x))-
 \partial_y-\mathcal{A}_{xj}(y)\bigl]\psi^{B}_{j}(x,y) &
    = \bigl[E-V_j(x,y)\bigl]\,\psi^{A}_j(x,y)\\
\bigl[-i(\partial_x+\mathcal{A}_{yj}(x))+\partial_y-
\mathcal{A}_{xj}(y)\bigl]\,\psi^{A}_j(x,y) &
    = \bigl[E-V_j(x,y)\bigl]\,\psi^{B}_j(x,y)
  \,.
\end{align}
\end{subequations}
 Taking into account the potential profile, 
 translation invariance
 in the $y$-direction and using (\ref{eq:01}), we write \eqref{eq:0001}
 as 
 \begin{subequations}\label{eq:0002}
\begin{align}
 -i\bigl[(\partial_x-\delta t_j)+ k_y\bigl]\varphi^{B}_j(x)e^{\textbf{\emph{i}}k_y y} &
    = \bigl[E-V_j(x)\bigl]\,\varphi^{A}_j(x)e^{\textbf{\emph{i}}k_y y}\\
-i\bigl[(\partial_x+\delta t_j)- k_y\bigl]\,\varphi^{A}_j(x)e^{\textbf{\emph{i}}k_y y} &
    = \bigl[E-V_j(x)\bigl]\,\varphi^{B}_j(x)e^{\textbf{\emph{i}}k_y y}
\end{align}
\end{subequations}
with $\psi^{A/B}_{j}(x,y)=\varphi^{A/B}_j(x)e^{\textbf{\emph{i}}k_y y}$.
It is convenient to introduce in $j$-th region the dimensionless quantities
$\mathbb{V}_j=V_j/E_F$, $\varepsilon_j=E_j/E_F$, $\delta\tau_j=\delta t_j/E_F$ with $E_F=\hbar v_F/d$.
Thus, the solutions in each $j$-region are
given by%
\begin{eqnarray}\lb{psx}
  \psi_{j}(x,y) = \psi_j(x)e^{\textbf{\emph{i}}k_y y} =
    w_j(x)D_j e^{\textbf{\emph{i}}k_y y}
\end{eqnarray}
where the two matrices read as
\begin{equation}\label{eq:0003}
        w_{j}(x)=
        \left(%
                \begin{array}{cc}
                        e^{ \textbf{\emph{i}} k_{j} x} & e^{ -\textbf{\emph{i}} k_{j} x} \\
                        s_{j} z_{j} e^{ \textbf{\emph{i}} k_{j} x}  & -s_{j}
                        z_{j}^{-1} e^{- \textbf{\emph{i}} k_{j} x} \\
                \end{array}
        \right),\qquad D_j=
        \left(%
                \begin{array}{c}
                        \alpha_j \\
                        \beta_j \\
                \end{array}
        \right)
\end{equation}
$\psi_j(x)$ is a spinor of components $\varphi^{A}_j(x)$ and  $\varphi^{B}_j(x)$,
$\alpha_j$ and $\beta_j$ being the amplitudes of positive and
negative propagation wave functions inside the $j$-th region, respectively. We show that
the associated eigenvalues are
\begin{equation}\label{Naf}
  \varepsilon-\mathbb{V}_j
    = s_j\sqrt{(k_j d) ^{2}+ \left(k_yd- \delta\tau_j \right)^2}
\end{equation}
where $ s_j=\text{sign}(\varepsilon-\mathbb{V}_j)$ is the usual sign function. It is clearly seen that
in the energy spectrum of
the $j$-th strained region \eqref{Naf}, the component $k_yd$ of the wave vector is shifted by $\delta\tau_j$
compared to that of the pristine graphene. This behavior was already encountered in our previous  work \cite{Bahlouli}
where we had a magnetic field that also shifted the wave vector  $ k_y $ by $ \frac{d}{l_{B}^{2}} $, $l_{B}=1/\sqrt{B_0}$
is the magnetic length and  $B_0$ is the strength of the magnetic field.
Hence from this behavior we can say that the deformation (strain effect) behaves like an effective magnetic field,
and the two shifts play the role of a mass term.

\section{Transmission and conductance}

To determine the transmission probability, we use
the boundary conditions  applied successively at the  interfaces  along the $x$-direction
and  evaluate the current densities in the incident, reflected and transmitted regions.
Indeed,
at interfaces $x_j=j \varrho$,
we obtain the transfer matrix
$
D_j=M_jD_{j+1}
$
with the matrix
\begin{equation}\label{eq:0004}
        M_{j}=
        \left(%
                \begin{array}{cc}
                        e^{ \textbf{\emph{i}} k_{j} j \varrho} & e^{ -\textbf{\emph{i}} k_{j} j \varrho} \\
                        s_{j} z_{j} e^{ \textbf{\emph{i}} k_{j} j \varrho}  & -s_{j}
                        z_{j}^{-1} e^{- \textbf{\emph{i}} k_{j} j \varrho} \\
                \end{array}
        \right)^{-1} \left(%
                \begin{array}{cc}
                        e^{ \textbf{\emph{i}} k_{j+1} j \varrho} & e^{ -\textbf{\emph{i}} k_{j+1} j \varrho} \\
                        s_{j+1} z_{j+1} e^{ \textbf{\emph{i}} k_{j+1} j \varrho}  & -s_{j+1}
                        z_{j+1}^{-1} e^{- \textbf{\emph{i}} k_{j+1} j \varrho} \\
                \end{array}
        \right)
\end{equation}
and we have
\begin{eqnarray}\label{eq:0005}
     k_j&
    = \frac{1}{d}\sqrt{\left(\varepsilon-\mathbb{V}_j\right)^2-\left(k_yd- \delta\tau_j\right)^2}, \qquad
z_j&
    = \frac{k_j-i\left(k_y-\frac{\delta \tau_j}{d}\right)}{\sqrt{k_j^{2}+\left(k_y-\frac{\delta \tau_j}{d}\right)^{2}}}.
\end{eqnarray}
with
$d=3\varrho$, $s_1=s_5$, $k_1=k_5$, $z_1=z_5$, $\delta \tau_j
=(0,\delta \tau_2,0,\delta \tau_4,0)$
and $\mathbb{V}_j 
=(0,\mathbb{V}_2,0,\mathbb{V}_4,0)$.
 After some lengthy algebras,  we
obtain
the transfer matrix
\begin{equation}\label{eq:0006}
  M=M_1 M_2 M_3 M_4=
  \left(\begin{matrix}
         M_{11} & M_{12}\\
         M_{11} & M_{12}
        \end{matrix}\right)
\end{equation}
Since the incident and reflected amplitudes
are defined by \beq
D_{1}=
        \left(%
                \begin{array}{c}
                        1 \\
                        r \\
                \end{array}%
    \right), \qquad  D_{5}=
    \left(%
                \begin{array}{c}
                        t \\
                        0 \\
                \end{array}%
        \right)
        \eeq
        then
the wavefunctions in the \emph{input} and \emph{output} regions are connected by the matrix $M$ as
\begin{equation}\label{eq:0007}
        D_{1}=M D_{5}.
\end{equation}
On the other hand,
the eigenspinors $ \psi_j(x)$, appearing in \eqref{psx}, can be decomposed  as
\begin{equation}\label{eq:0008}
  \psi_j(x)=\alpha_j  \psi_j^{+}(x)+ \beta_j \psi_j^{-}(x)
\end{equation}
where the two components are given by
\begin{eqnarray}\label{eq:0009}
  \psi_j^{+}(x) = \left(
                       \begin{array}{c}
                         1 \\
                         s_j z_j \\
                       \end{array}
                     \right) e^{ \textbf{\emph{i}} k_{j}x}, \qquad
  \psi_j^{-}(x) = \left(
                       \begin{array}{c}
                         1 \\
                         -s_j z_j^{-1} \\
                       \end{array}
                     \right) e^{- \textbf{\emph{i}} k_{j}x}
\end{eqnarray}
which are
the positive and negative propagating
spinors associated to their amplitudes $\alpha_j$ and $\beta_j$, respectively.
In \emph{input} region, conservation of the mode $k_y$ gives the relation 
\begin{equation}\label{0016}
{ \sin\theta_j = \frac{1}{|\varepsilon-\mathbb{V}_j|}\left(\varepsilon\sin\theta -  {\delta \tau_j}\right)}
\end{equation}
where {$\theta_j$ is the angle of propagation  of Dirac fermions within $j$-region},
$\theta=\theta_1$ 
is the incident angle and $\varepsilon$ is the incident energy.
{From \eqref{0016} we notice that  
the strain directly affects the transmission angle of Dirac fermions \cite{AIP}.}

At this stage, we can introduce the current density corresponding to
our system. Indeed, we show that
the incident, reflected and transmitted
current densities can be written as
\begin{subequations}\label{eq:0010}
\begin{align}
  &J_x^{\sf in}  ={\psi_1^+(x)}^{\dag}\sigma_x{\psi_1^+(x)} 
  =
  2s_1\frac{k_1}{\sqrt{k_1^{2}+k_y^{2}}}
\\
  & J_x^{\sf re}  ={\psi_1^-(x)}^{\dag}\sigma_x{\psi_1^-(x)} 
  =2s_1|r|^{2}\frac{k_1}{\sqrt{k_1^{2}+k_y^{2}}}
\\
  & J_x^{\sf tr}  ={\psi_5^+(x)}^{\dag}\sigma_x{\psi_5^+(x)} 
  =2s_5|t|^{2}\frac{k_5}{\sqrt{k_5^{2}+k_y^{2}}}
\end{align}
\end{subequations}
giving rise to the transmission and reflection probabilities
\begin{eqnarray}\label{eq:0011}
T =\frac{\left|J_x^{\sf tr} \right|}{\left|J_{x}^{\sf in} \right|}
=\frac{s_5 k_5}{s_1k_1}\frac{\sqrt{k_1^{2}+k_y^{2}}}{\sqrt{k_5^{2}+k_y^{2}}}|t|^2, \qquad
R =\frac{\left|J_{x}^{\sf re} \right|}{\left|J_{x}^{\sf in} \right|} =|r|^2.
\end{eqnarray}
Due to the symmetry of the double barrier configuration in the \emph{input} and \emph{output} regions,
we have the relations $s_{1}=s_{5}$ and $k_1 = k_5$. Thus, using \eqref{eq:0007} to obtain 
\begin{equation}\label{0015}
  T = |t|^{2}
  = \frac{1}{|M_{11}|^{2}}.
\end{equation}

Since the transmission probability $T$ is determined for  each mode $k_y$, then
we derive the conductance $G$ at zero temperature. Indeed, using the definition
\cite{Buttiker1985} we find the conductance in the unit systems
\begin{eqnarray}
        G&=& \frac{2e^2}{\pi} \int_{-E}^{E}T_n(E,k_y)\frac{dk_y}{2\pi/L_y}\\
        &=&G_0\int_{-\pi/2}^{\pi/2}T_n(\varepsilon,\theta)\cos\theta d\theta\label{Conductance}
\end{eqnarray}
where $L_y$ is the sample size along the $y$-direction and $G_0=\frac{e^2\varepsilon L_y}{\pi^2 d}$
is the unit conductance, $d$ is
 the total barrier width. These results will numerically be investigated to analyze
the system behavior and
underline its basic features.

\section{Results and discussions}

In the beginning we analyze only the double strain effects on
the graphene systems, which means that we keep strain parameter
 in two regions ($\textbf{\textcircled{2}}$, $\textbf{\textcircled{4}}$)
 and forget about the double barrier potential $\mathbb{V}_2=\mathbb{V}_4=0$. 
 Note that, the conservation  imposes $k_1=k_3=k_5$, $s_1=s_3=s_5$, which will be considered
 in the forthcoming analysis.
 
\begin{figure}[!ht]\centering
    \subfloat[]{
        \includegraphics[scale=0.18]{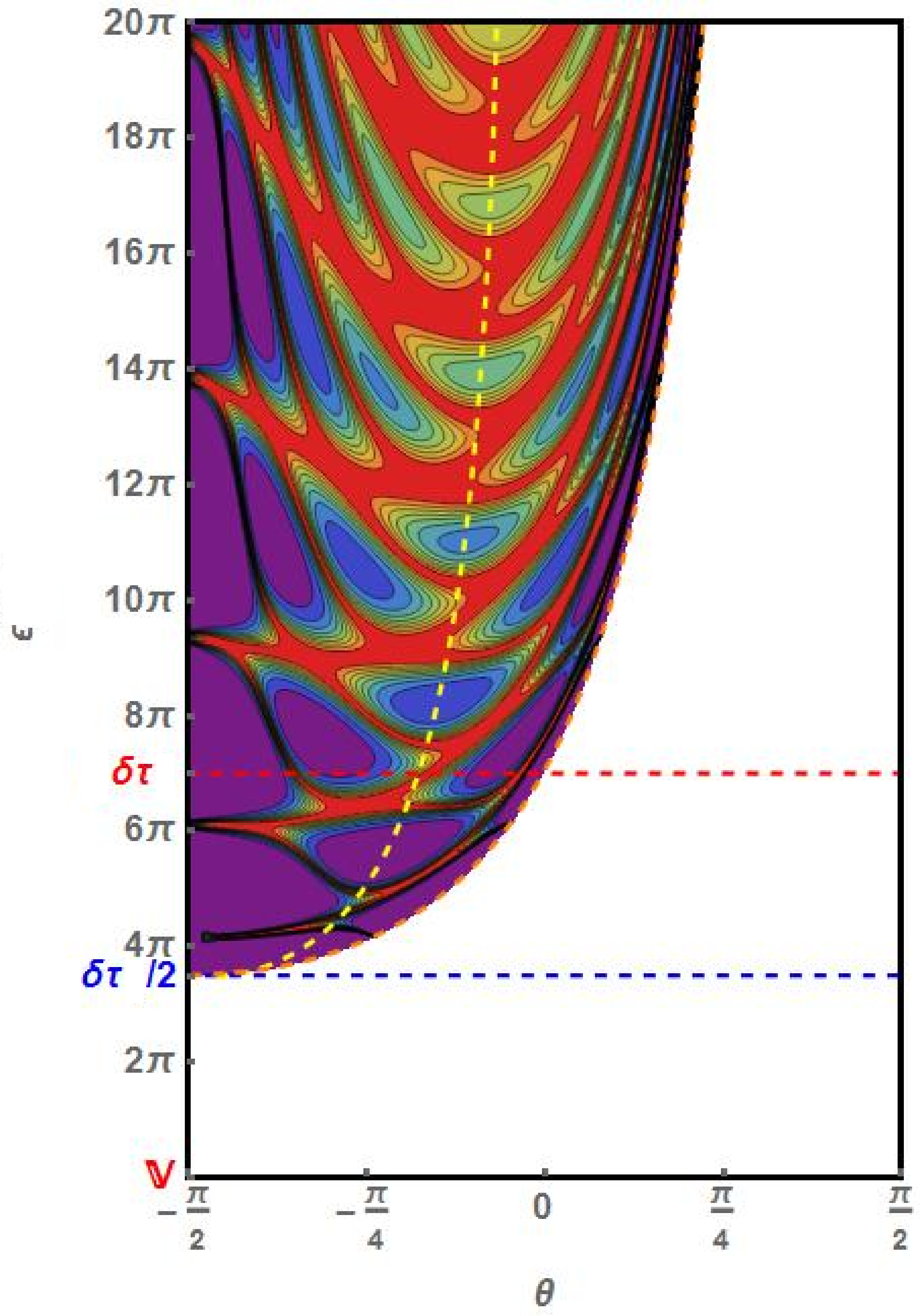}
        \label{fig01:SubFigA}
    }\hspace{-0.2cm}
    \subfloat[]{
        \includegraphics[scale=0.18]{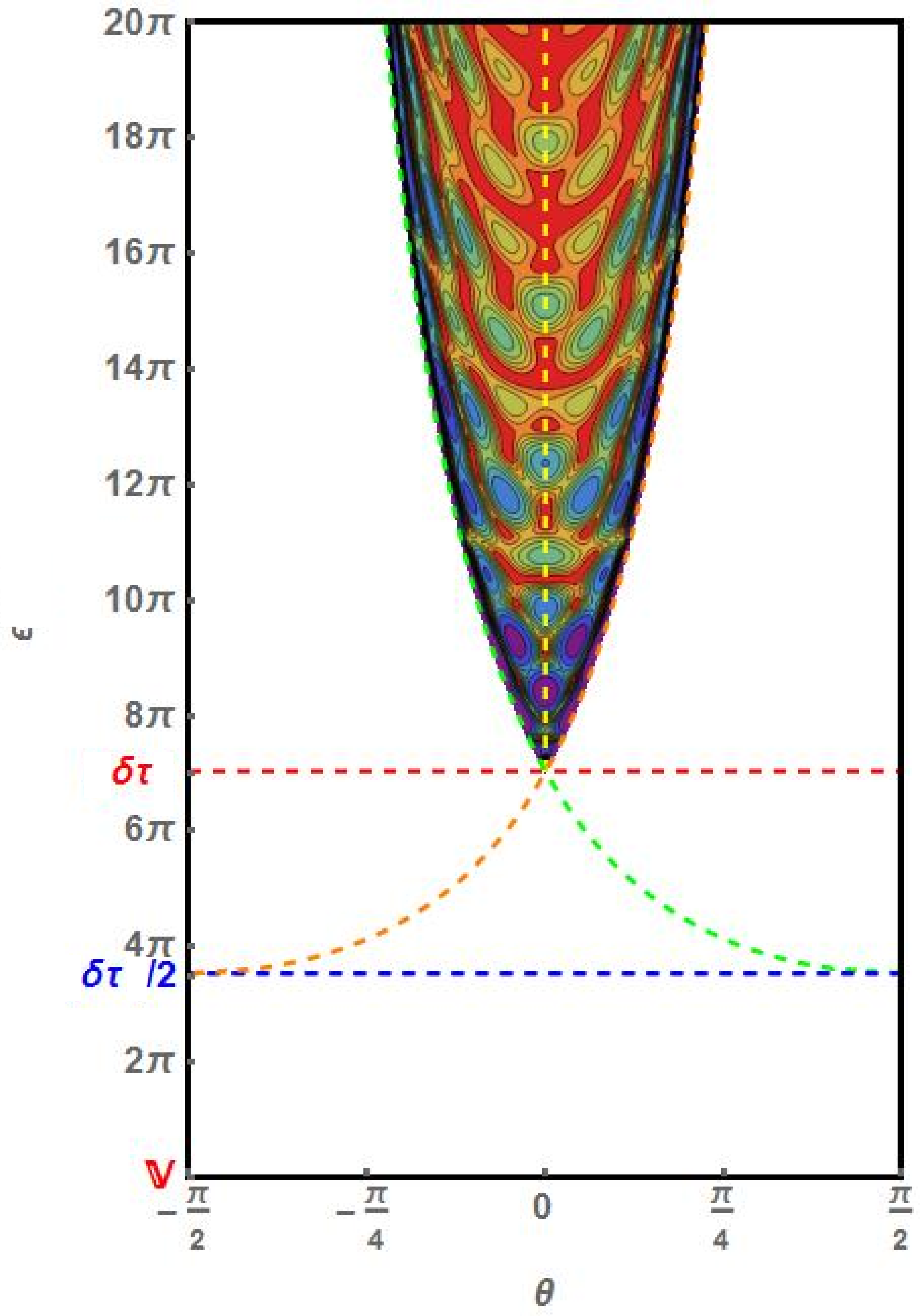}
        \label{fig01:SubFigB}
    }\hspace{-0.2cm}
        \subfloat[]{
        \includegraphics[scale=0.18]{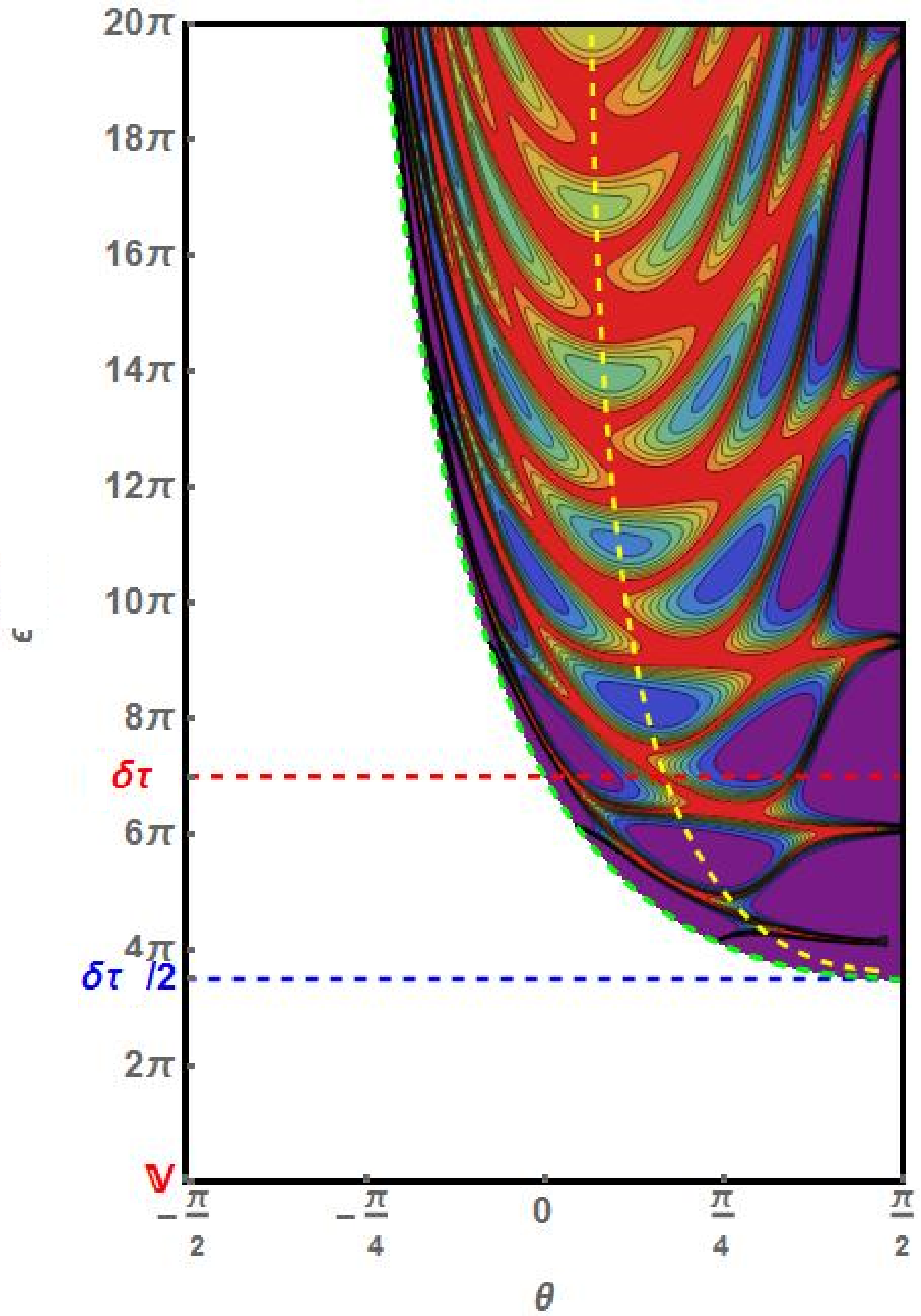}
        \includegraphics[scale=0.1955]{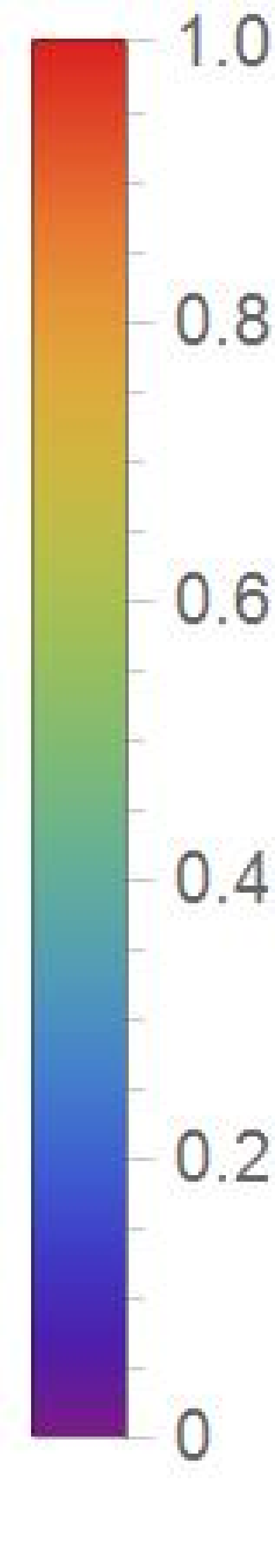}
        \label{fig01:SubFigC}
    }
    \caption{
                (color online) Density plot of transmission probability $T$ versus incident
                energy $\varepsilon$ and incident angle $\theta$ with $\mathbb{V}_2=\mathbb{V}_4=0$,
                $\varrho=300$, $d=3\varrho$, $\delta \tau=7\pi$ (red dashed line), $\frac{\delta \tau}{2}=\frac{7\pi}{2}$ (blue dashed line),
                ${{\varepsilon=\frac{\delta \tau}{1+\sin\theta}}}$ (green dashed line),
                ${\varepsilon=\frac{\delta \tau}{1-\sin\theta}}$ (orange dashed line)
                and the symmetry curve (yellow dashed line),  for \protect \subref{fig01:SubFigA}:
                $\delta \tau_2 =\delta \tau_4=-\delta \tau$, \protect \subref{fig01:SubFigB}:
                $\delta \tau_2 =-\delta \tau_4=\pm\delta \tau$,
                \protect \subref{fig01:SubFigC}: $\delta \tau_2 =\delta \tau_4=\delta \tau$.
        }
        \label{Fig01}
\end{figure}

\noindent
In Figure \ref{Fig01},
we present density plots of transmission probability as function of incident energy $\varepsilon$
and incident angle $\theta$.
%
We show that
the white
zones that are limited in Figure \ref{Fig01}\subref{fig01:SubFigA} 
by
${\varepsilon=\frac{\delta \tau}{1-\sin\theta}}$ (orange dashed line),
in Figure \ref{Fig01}\subref{fig01:SubFigB} 
by
${\varepsilon=\frac{\delta \tau}{1-\sin\theta}}$ and
${\varepsilon=\frac{\delta \tau}{1+\sin\theta}}$ (green dashed line),
and in Figure \ref{Fig01}\subref{fig01:SubFigC} 
by
${\varepsilon=\frac{\delta \tau}{1+\sin\theta}}$
represent forbidden zones (FZs). Otherwise, the FZs are determined by angles
${\theta =\arcsin \left(\frac{\pm(\varepsilon- \delta \tau) }{\varepsilon}\right)}$,
which give rise to a phenomenon resembling beam collimation in optics. These relations can be obtained either from Figure \ref{Fig02} (see below) or directly from
the conservation of the mode $k_y$ \eqref{0016}.
We notice that the tunneling effect is always completely suppressed:
in Figures (\ref{Fig01}\subref{fig01:SubFigA}, \ref{Fig01}\subref{fig01:SubFigC})
if $\varepsilon <\frac{\delta \tau}{2} $  (dashed line) 
which was found in
the case of a single barrier \cite{8}, and in Figure \ref{Fig01}\subref{fig01:SubFigB}
if $\varepsilon <\delta \tau$  (red dashed line) corresponding to the two collimations generated mutually by
contraction and compression strains. Figure \ref{Fig01} tells us that the
Klein paradox does not always exist at normal incidence angle $\theta=0$. Furthermore, 
the transmission is not symmetrical with respect to the normal incidence angle
in Figures (\ref{Fig01}\subref{fig01:SubFigA}, \ref{Fig01}\subref{fig01:SubFigC}),
but it restores its  symmetry in Figure \ref{Fig01}\subref{fig01:SubFigB}. Note that in Figure \ref{Fig01} the dashed yellow
line is a symmetry curve of  transmission. We observe that  in Figure \ref{Fig01}\subref{fig01:SubFigB}
there is a perfect symmetry, on the other in Figures (\ref{Fig01}\subref{fig01:SubFigA},
\ref{Fig01}\subref{fig01:SubFigC}) there is absence of symmetry. In Figures \ref{Fig01},
purple zones  represent transmission gaps ($T=0$) and red zones  represent total transmission ($T=1$)
i.e.  Klein tunneling effect. Inspecting the last Figures, exposes the following symmetry relations
\begin{subequations}
\begin{eqnarray}
  && T(\theta,\delta \tau ,\delta \tau)=T(-\theta ,-\delta \tau,-\delta \tau)\lb{24a}\\
 && T(\theta,\delta \tau,-\delta \tau)=T(\theta,-\delta \tau,\delta \tau)\lb{24b}\\
 && T(\theta,\pm\delta \tau,\mp\delta \tau)=T(-\theta,\pm\delta \tau,\mp\delta \tau).\lb{24c}
\end{eqnarray}
\end{subequations}

Collimation is due to filtering effect at certain incidence angles,  Figure \ref{Fig02}
shows this effect which is best understood by inspecting the phase-spaces \cite{8}. The Fermi surface of pristine graphene
is highlighted in yellow and has a dispersion given by $\varepsilon^{2}=d^{2}(k_1^{2}+k_y^{2})$.
The strained graphene, on the other hand, exposes a green and orange Fermi surface for compression and contraction strain
whose equations are, respectively,
\beq
\varepsilon^{2}=d^{2}k_{2,4}^{2}+(d k_y+\delta\tau)^{2},\qquad
\varepsilon^{2}=d^{2}k_{2,4}^{2}+(d k_y-\delta\tau)^{2}.
\eeq
Conservation of energy and momentum $k_y$ immediately leads to a sector of
allowed incident angles  (red surfaces) whose openings are limited by the
red arrows in Figure \ref{Fig02} and corresponding exactly to the allowed
incident angles shown  in Figure \ref{Fig01}. If $\delta \tau$ exceeds $2 \varepsilon$,
the red surface will be omitted and therefore tunneling in strained double barrier is
suppressed. Figures (\ref{Fig02}\subref{fig02:SubFigA}, \ref{Fig02}\subref{fig02:SubFigC})
show that the two red surfaces are symmetric with  respect to the normal
incidence angle but in Figure \ref{Fig02}\subref{fig02:SubFigB} the red surface is symmetric
with respect to the normal incidence angle. By combining Figures (\ref{Fig01}, \ref{Fig02})
we can explain easily the Dirac fermions like collimation in strained double barrier.

\begin{figure}[!ht]\centering
    \subfloat[]{
        \includegraphics[scale=0.2]{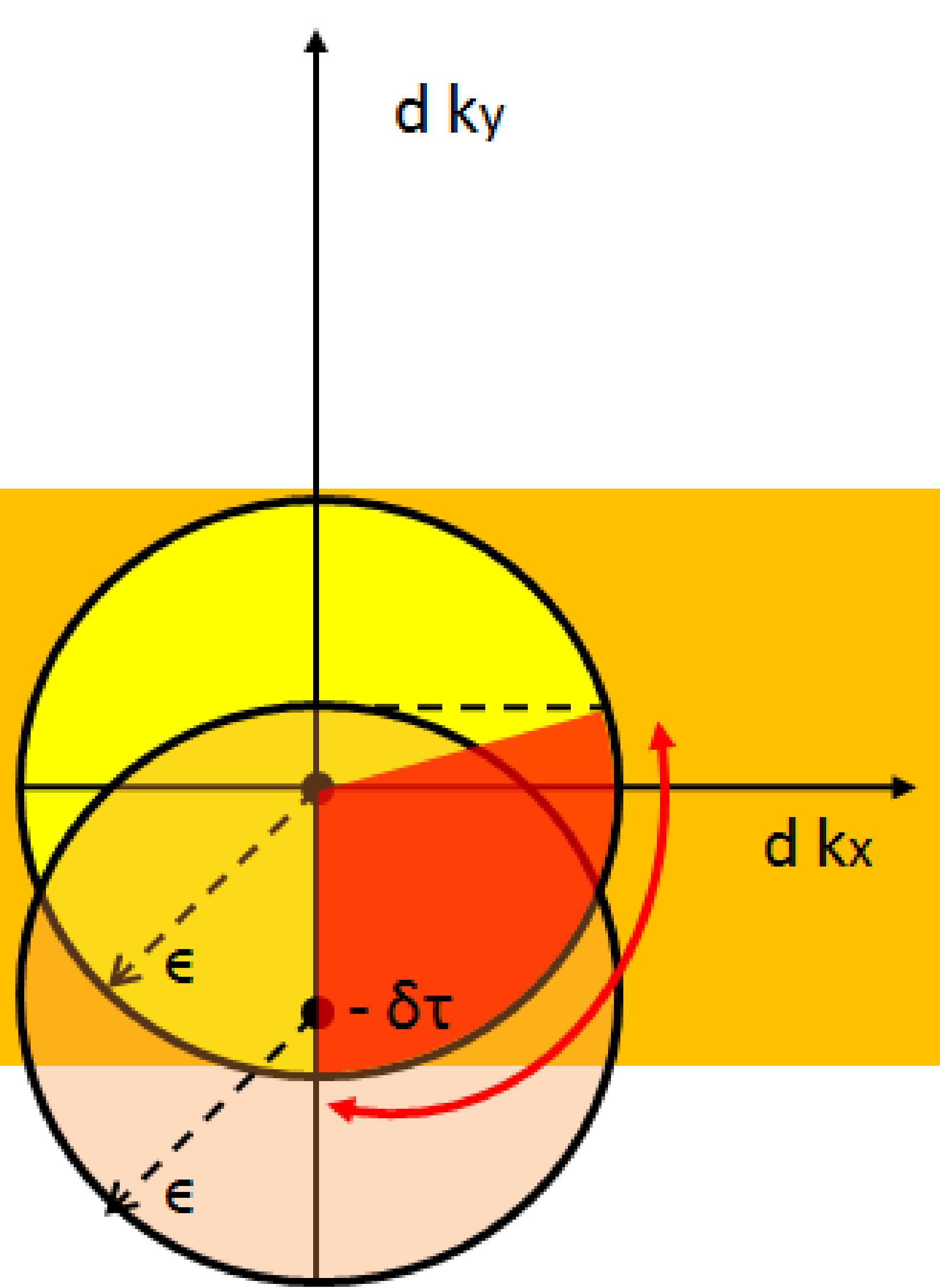}
        \label{fig02:SubFigA}\ \ \ \ \ \ \ \
    }\hspace{-0.02cm}
    \subfloat[]{
        \includegraphics[scale=0.2]{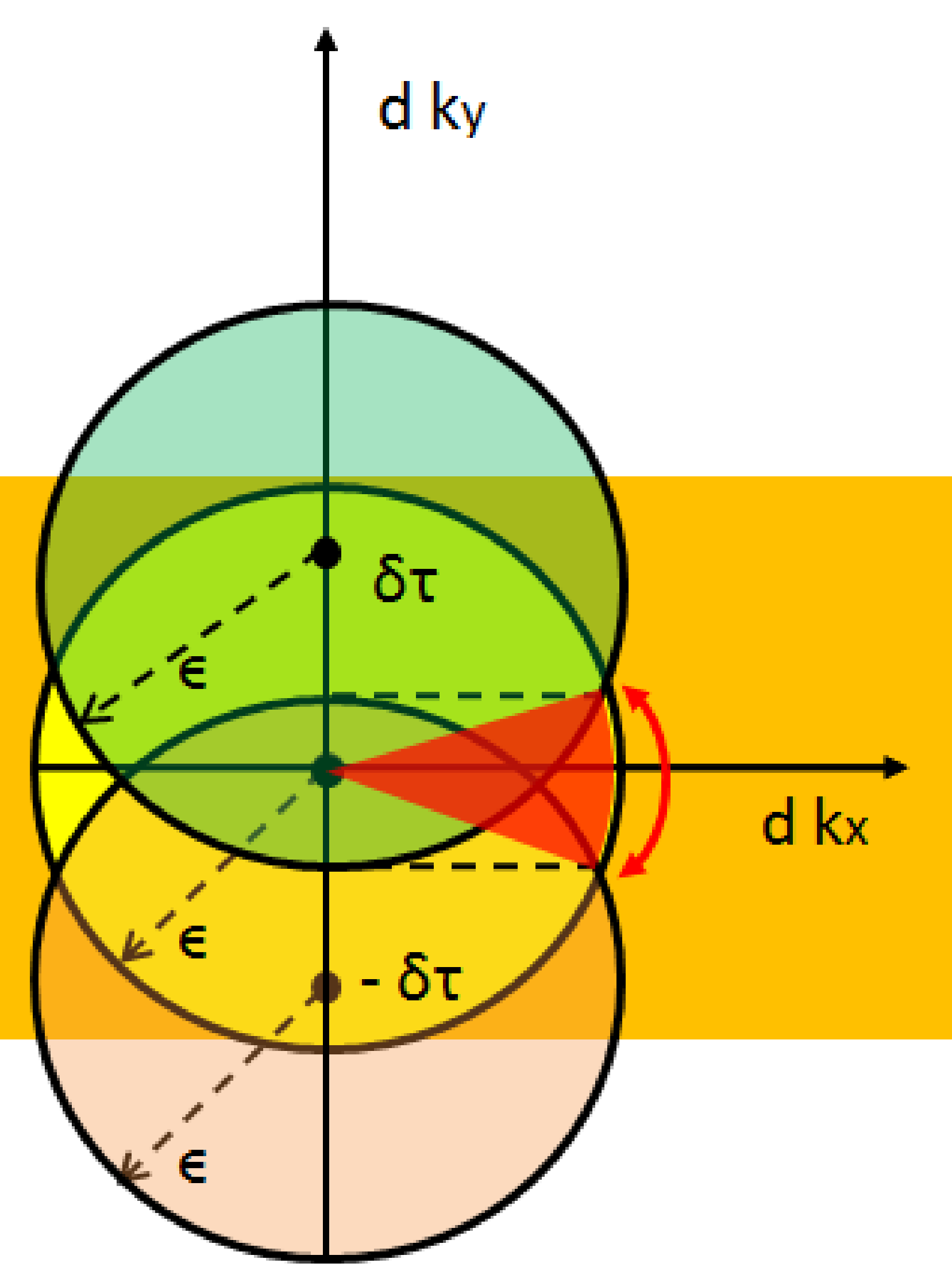}
        \label{fig02:SubFigB}\ \ \ \ \ \ \ \
    }\hspace{-0.02cm}
        \subfloat[]{
        \includegraphics[scale=0.2]{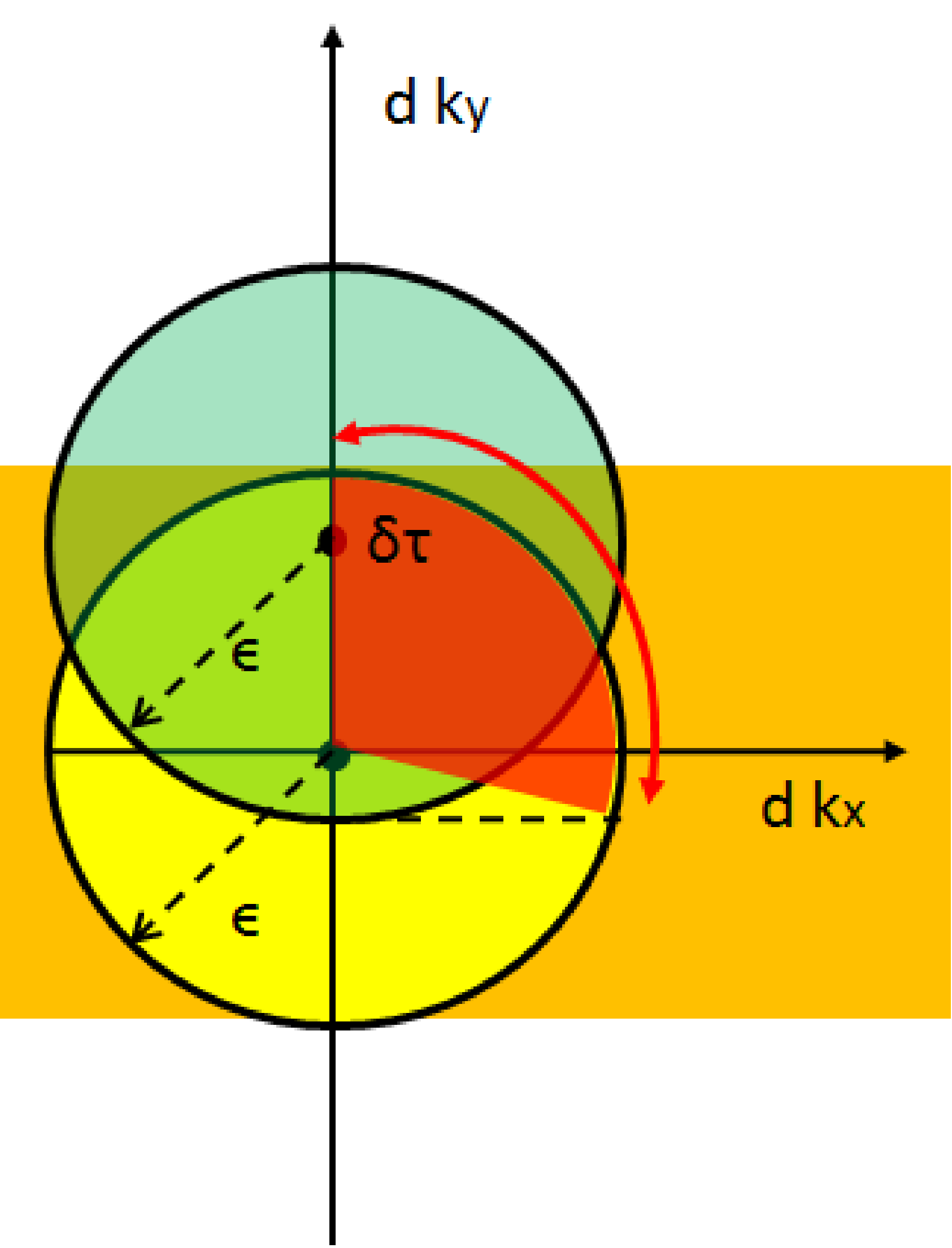}
        \label{fig02:SubFigC}
    }
    \caption{
                (color online) Fermi surfaces affected by strain in double barrier,  for \protect \subref{fig01:SubFigA}:
                $\delta \tau_2 =\delta \tau_4=-\delta \tau$, \protect \subref{fig01:SubFigB}:
                $\delta \tau_2 =-\delta \tau_4=\pm\delta \tau$,
                \protect \subref{fig01:SubFigC}: $\delta \tau_2 =\delta \tau_4=\delta \tau$.
        }
        \label{Fig02}
\end{figure}

\begin{figure}[!ht]\centering
    \subfloat[]{
        \includegraphics[scale=0.18]{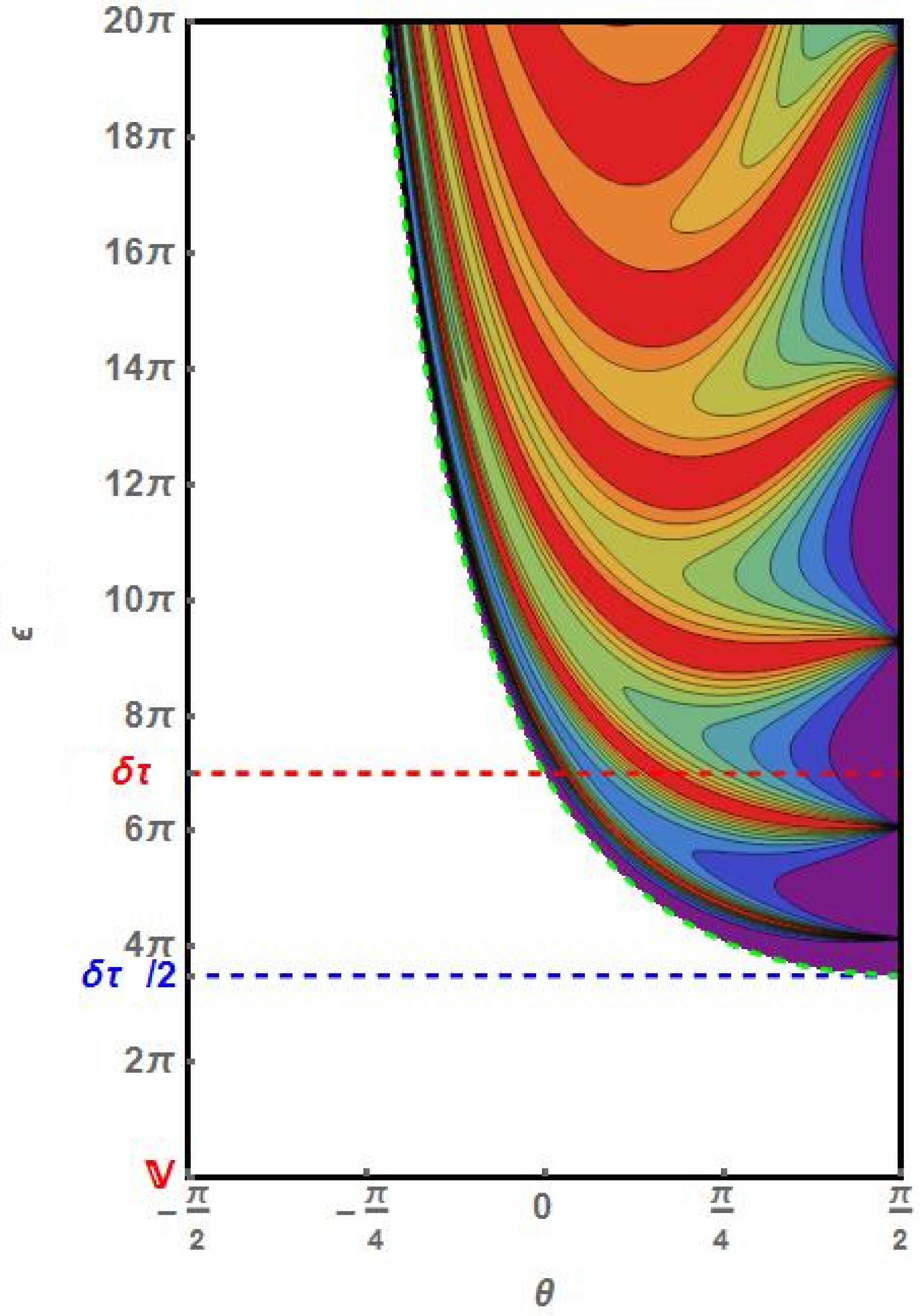}
        \label{fig03:SubFigA} \ \ \ \ \ \ \ \
    }\hspace{-0.01cm}
        \subfloat[]{
        \includegraphics[scale=0.18]{nouj02}
        \includegraphics[scale=0.1955]{0000}
        \label{fig03:SubFigC}
    }
    \caption{
                (color online) Density plot of transmission probability $T$ versus incident energy
                $\varepsilon$ and incident angle $\theta$ with $\mathbb{V}_2=\mathbb{V}_4=0$,
                $\varrho=300$, $d=3\varrho$, $\delta \tau=7\pi$ (red dashed line), $\frac{\delta \tau}{2}=\frac{7\pi}{2}$ (blue dashed line), ${\varepsilon=\frac{\delta \tau}{1+\sin\theta}}$
                 (green dashed line)
                and the symmetry curve (yellow dashed line).   \protect \subref{fig03:SubFigA}:
                Strained single barrier with $\delta \tau_2=\delta \tau, \delta \tau_4=0$.
                \protect \subref{fig03:SubFigC}: Strained double barrier with
                $\delta \tau_2=\delta \tau_4=\delta \tau=0$.
        }
        \label{Fig03}
\end{figure}

Considering the symmetry between $T(\theta,\delta \tau,\delta \tau)$ and $T(\theta,-\delta \tau,-\delta \tau)$
with respect to $\theta=0$, we will only study ($T(\theta,\delta \tau,\delta \tau)$, $T(\theta,\delta \tau,0) $)
and deduce the other symmetrical cases. In Figure \ref{Fig03}, 
we evaluate the transmission difference between a strained single barrier with
($\delta \tau_2=\delta \tau, \delta \tau_4=0$) and the  strained
double barrier with ($\delta \tau_2=\delta \tau_4=\delta \tau$), the two Figures
(\ref{Fig03}\subref{fig03:SubFigA}, \ref{Fig03}\subref{fig03:SubFigC}) have the same white forbidden
zone and therefore the same allowed incident angles. The double barrier introduces a radical
change as compared to in single barrier transmission. In Figure \ref{Fig03}\subref{fig03:SubFigA} the single barrier is
not symmetrical (no yellow curve),  it possess  separate total energy bands
(red color) having parabolic form that start with peaks at grazing incidence $\theta= \frac{\pi}{2}$ for $\delta\tau$
($\theta=-\frac{\pi}{2}$ for $-\delta\tau$), 
these peaks are intercalated by transmission gaps (purple zones). The first transmission gap starts
its location between energy $\frac{\delta \tau}{2}$ (blue dashed line)  and first peak, it coats the transmission density
plot on the green dashed line side ${\varepsilon=\frac{\delta \tau}{1+\sin\theta}}$
and becomes thin when the energy increases. Transmission gaps between peaks decreases
in width (compared to $\theta$) and increases in height (compared to $\varepsilon$) when the energy
increases. The  double barrier effect  is explained in Figure \ref{Fig03}\subref{fig03:SubFigC},
the density plot has a symmetry deformed with
respect to the yellow line. Indeed the total transmission bands (red zones) become fragmented and
contain islands of different transmission values, transmission gaps intercalated between the peaks
(at the grazing angle) are duplicated and  asymmetric with respect to the yellow
line, and the peaks are duplicated and asymmetric as well.
We observe that the energy
$\varepsilon=\frac{\delta \tau}{2}$ plays the role of mass term whose lower energy is forbidden.
In the presence of a mass term comes the first transmission gap followed by the first peak.

\begin{figure}[!ht]\centering
    \subfloat[]{
        \includegraphics[scale=0.336]{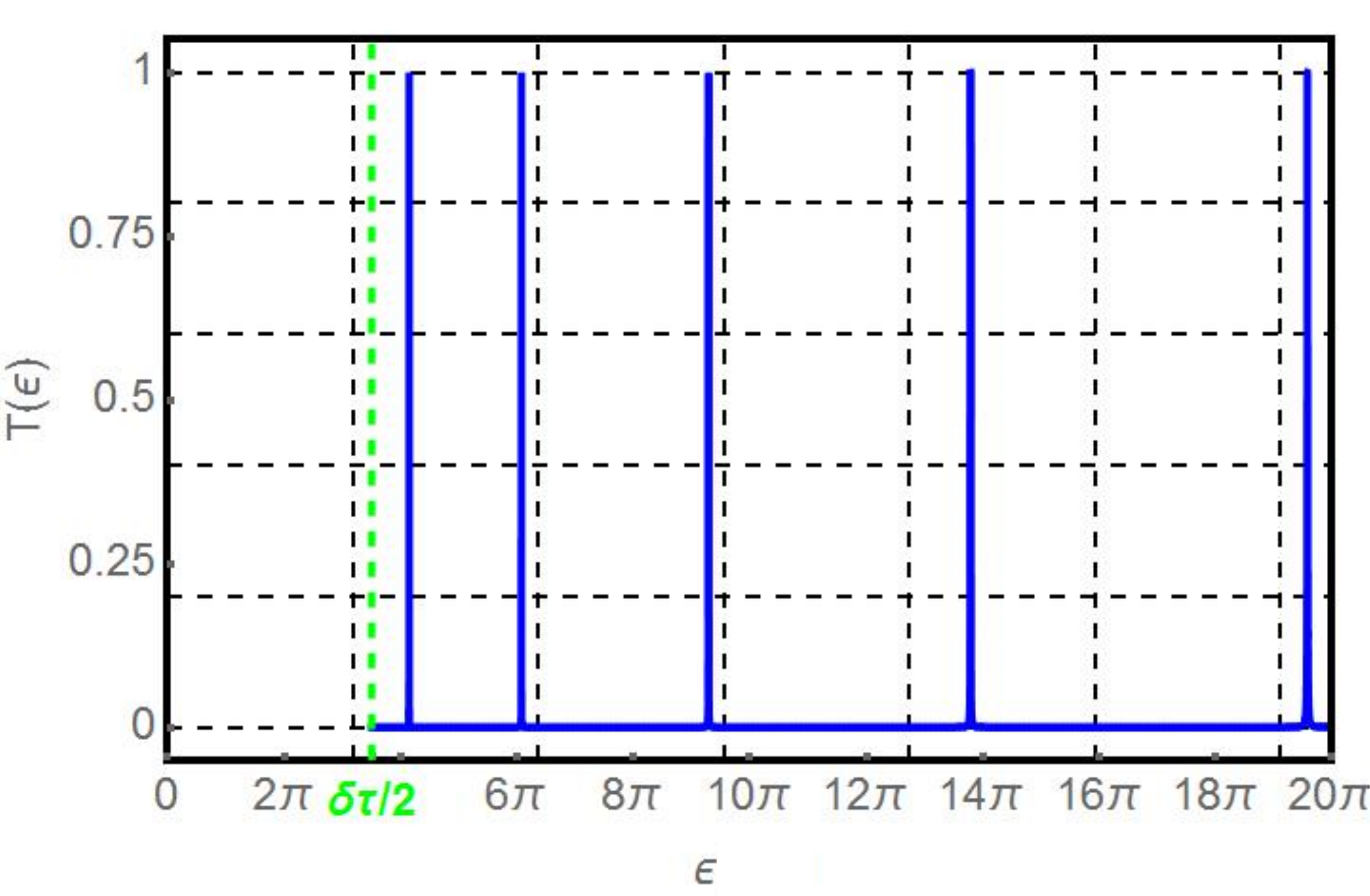}
        \label{fig04:SubFigA} \ \ \ \ \ \
    }\hspace{-0.2cm}
        \subfloat[]{
        \includegraphics[scale=0.336]{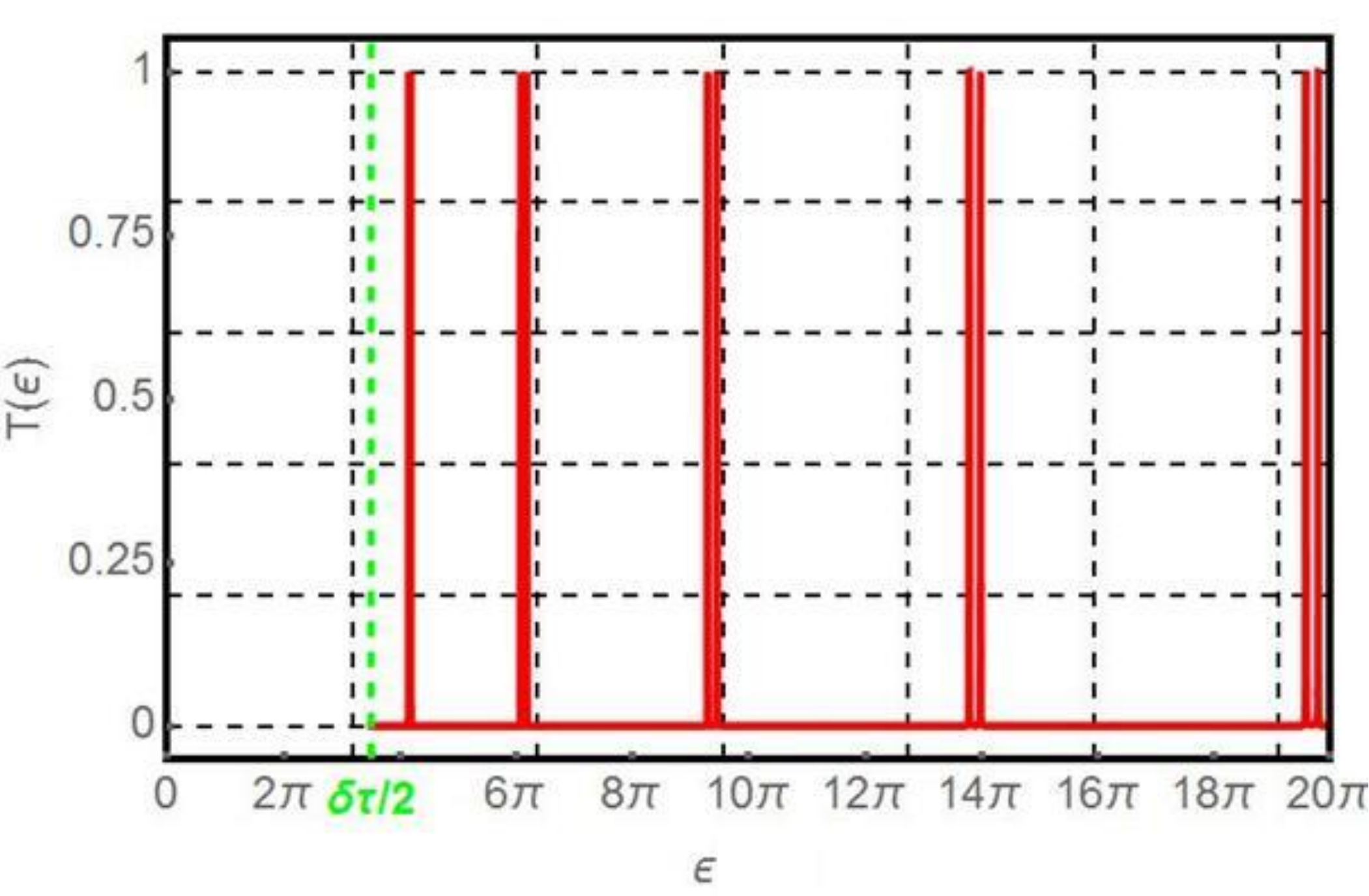}
        \label{fig04:SubFigC}
    }
    \caption{
                (color online) Transmission probability $T$ versus incident energy $\varepsilon$
                at grazing incidence $\theta=\pm \frac{\pi}{2}$ with $\mathbb{V}_2=\mathbb{V}_4=0$,
                $\varrho=300$, $d=3\varrho$, $\delta \tau=7\pi$ (green dashed line).
                \protect \subref{fig03:SubFigA}:
                Strained single barrier with $\delta \tau_2=\pm\delta \tau, \delta \tau_3=\delta \tau_4=0$.
                \protect \subref{fig03:SubFigC}: Strained double barrier with $\delta \tau_2=\pm\delta
                \tau_4=\pm\delta \tau$.
        }
        \label{Fig04}
\end{figure}

{To show difference between peaks behavior of the transmissions 
 in strained single  and 
 double barriers  at the grazing incidence angle $\theta=\frac{\pi}{2}$ for $\delta\tau$
($\theta=-\frac{\pi}{2}$ for $-\delta\tau$), we present Figure  \ref{Fig04}
with the same conditions as in  Figure \ref{Fig03}}.
In Figure \ref{Fig04}\subref{fig04:SubFigA}, the transmission probability $T$ versus
incident energy $\varepsilon $ at grazing incidence $\theta$ contains a series
of peaks intercalated with transmission gaps that increase as $\varepsilon $ increases.
In Figure \ref{Fig04}\subref{fig04:SubFigC}, $T$  behaves
like the one in the Figure \ref{Fig04}\subref{fig04:SubFigA} but with double resonance peaks,
the separation between peak doublet increases also with energy. Such
resonance peaks can  easily be explained by the doubling of the resonator
model of  Fabry-Perot in graphene \cite{zah1,zah2,zah3}.
The two barrier types (single and double)
have no effect on the width of the transmission gaps between  transmission peaks.
Figure \ref{Fig04}
shows also the forbidden transmission zone  between the zero energy $\varepsilon=0 $ and the mass-like term $\frac{\delta \tau}{2}$.

\begin{figure}[!ht]\centering
    \subfloat[]{
        \includegraphics[scale=0.08]{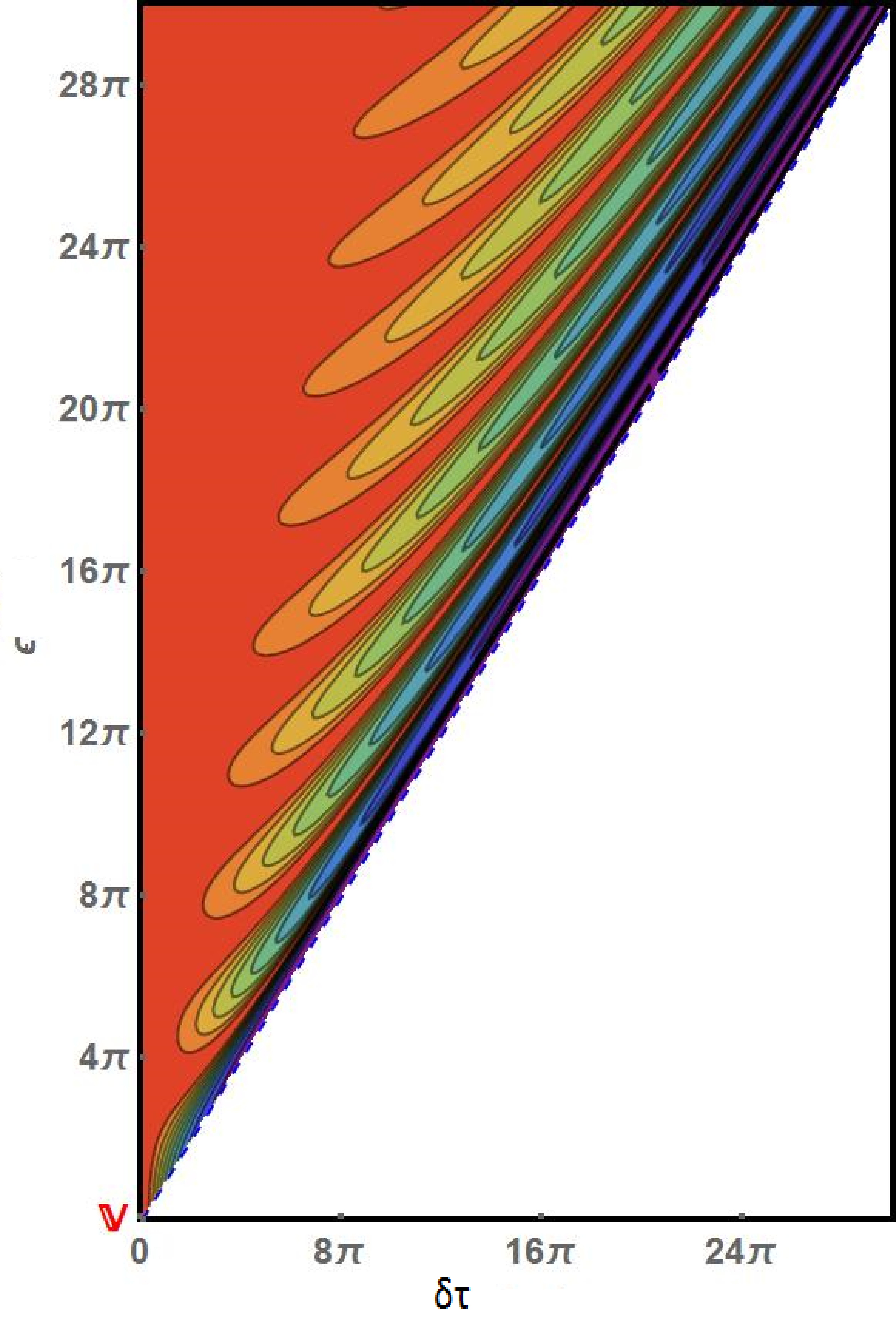}
        \label{fig05:SubFigA} \ \ \ \
    }\hspace{-0.3cm}
    \subfloat[]{
        \includegraphics[scale=0.08]{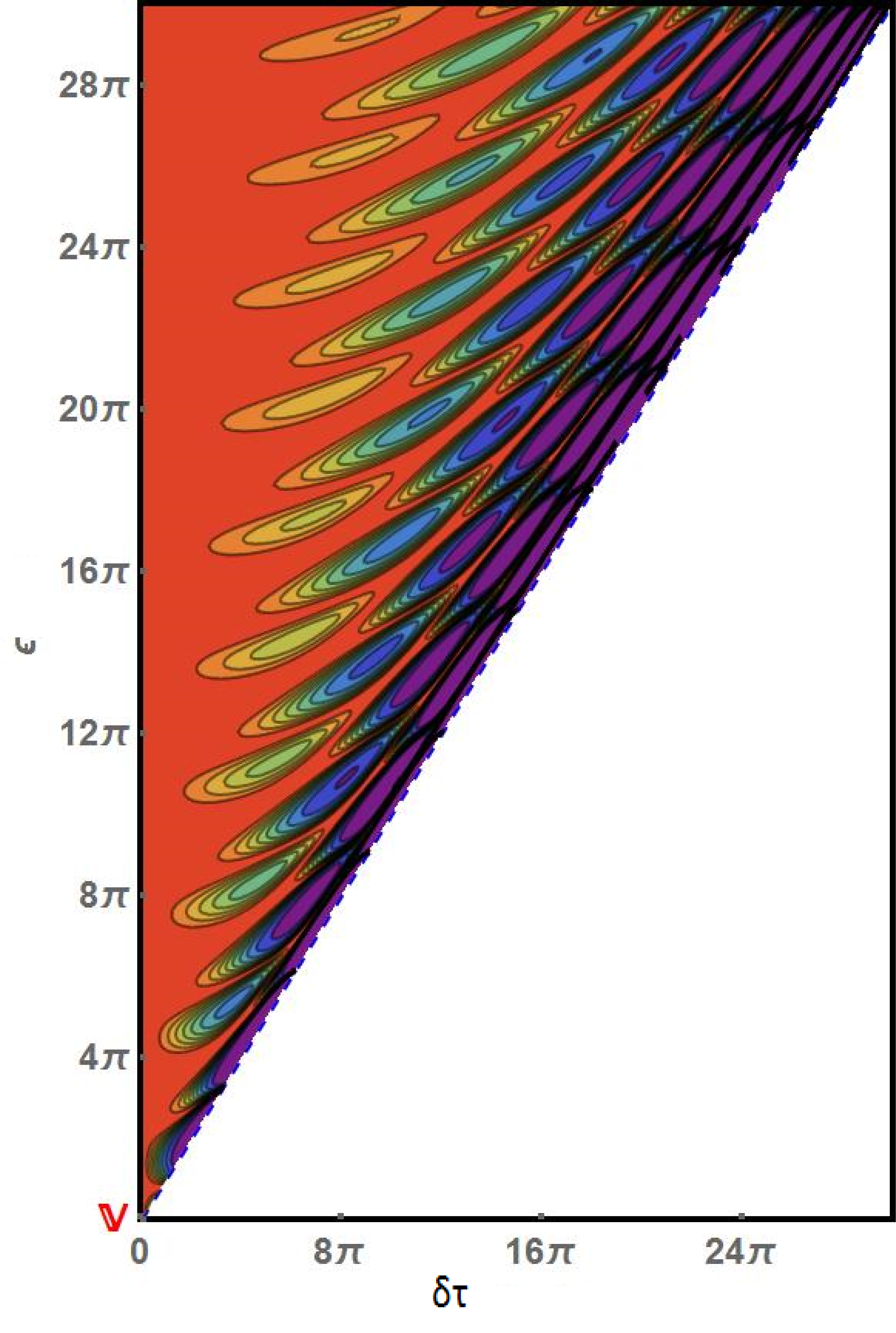}
        \label{fig05:SubFigB} \ \ \ \
    }\hspace{-0.3cm}
        \subfloat[]{
        \includegraphics[scale=0.08]{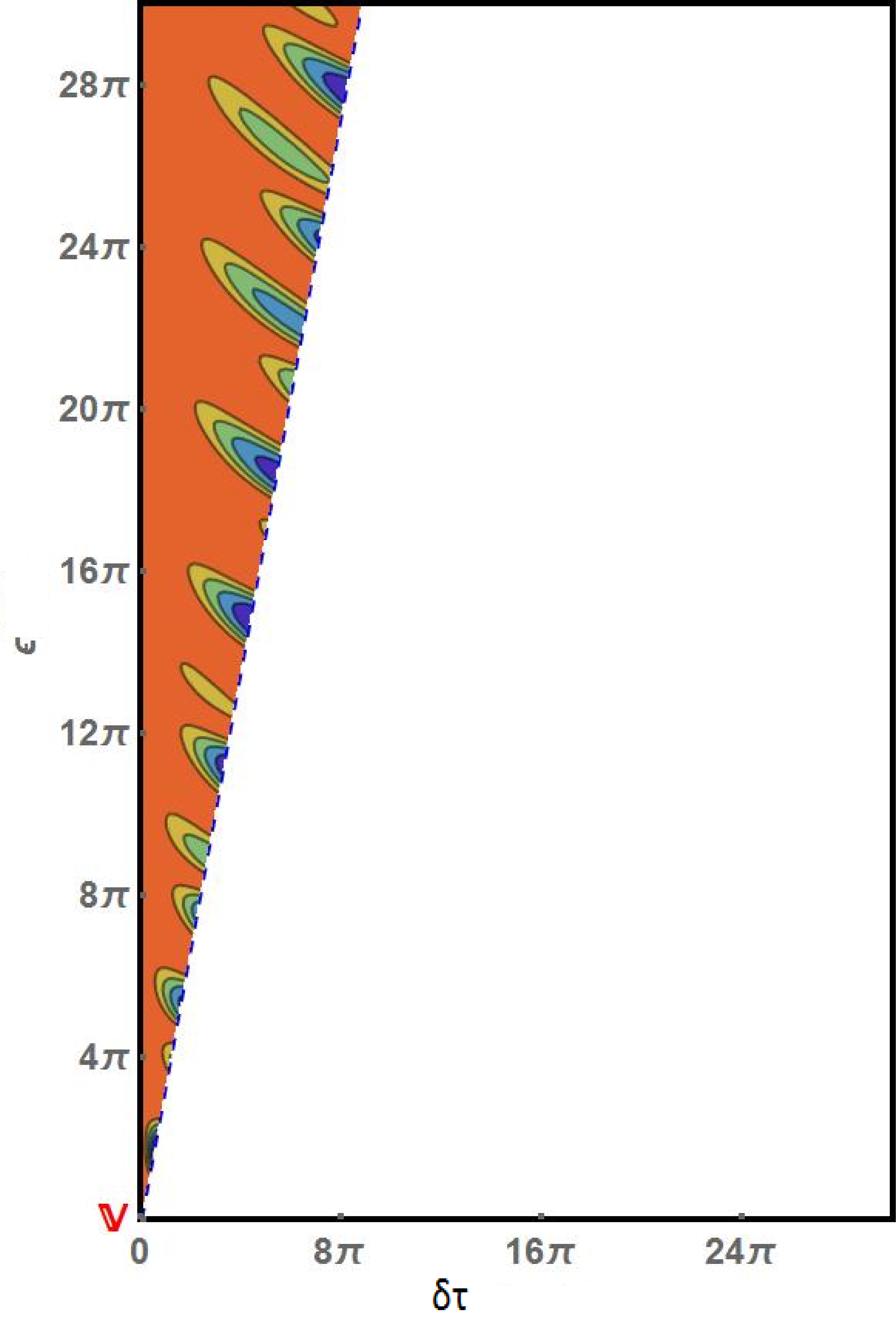}
        \includegraphics[scale=0.1555]{0000}
        \label{fig05:SubFigC}
    }
    \caption{
                (color online) Density plot of transmission probability $T$ versus incident
                energy $\varepsilon$ and  $\delta \tau$ with $\mathbb{V}_2=\mathbb{V}_4=0$,
                $\varrho=300$, $d=3\varrho$.  \protect \subref{fig05:SubFigA}: Single barrier
                transmission density plot in normal incidence with
                $\delta \tau_2 =\pm\delta \tau$, $\delta \tau_4 =0$ ) or $\delta \tau_2 =0$,
                $\delta \tau_4 =\pm\delta \tau$. \protect \subref{fig05:SubFigB}:  Double barrier
                transmission density plot in normal incidence with $\delta \tau_2 =-\delta \tau_4=\pm\delta \tau$.
                \protect \subref{fig05:SubFigC}: Double barrier  transmission density plot in
                incident angle $\theta= \frac{\pi}{4}$ with $\delta \tau_2 =-\delta \tau_4=\pm\delta \tau$.
        }
        \label{Fig05}
\end{figure}

Figure \ref{Fig05} shows
the transmission density plot versus incident energy $\varepsilon$ and
deformation $\delta \tau$ where
white zones correspond to forbidden transmission, purple zones correspond to transmission
gaps and red zones correspond to total transmission. The allowed transmission corresponds
to the energies ${\varepsilon\geq\frac{\delta \tau}{1-\sin\theta}}$.
In Figures (\ref{Fig05}\subref{fig05:SubFigA}, \ref{Fig05}\subref{fig05:SubFigB}),
 the single   and double barrier transmissions are illustrated respectively at normal
incidence $\theta=0$, we observe  that
near $\delta \tau=0$ (i.e. the pristine graphene is slightly deformed)
always there is the effect of Klein paradox (red color). As long as the deformation increases,
the different  transmission values  appear and we end up with  transmission gaps just
before the energy ${\varepsilon=\frac{\delta \tau}{1-\sin\theta}}$.
The contribution of the double barrier compared to the single one is marked by the appearance of
the islands of transmissions, 
which are surrounded by  total transmission zones
and  multiplication of transmission gap zones on the sides of
${\varepsilon=\frac{\delta \tau}{1-\sin\theta}}$.
The number of these islands and transmission gaps increase as long as  energy increases.
Figure \ref{Fig05}\subref{fig05:SubFigC} $(\theta=\frac{\pi}{4})$ shows that if the angle
of incidence increases the forbidden zone increases, islands and transmission gaps decrease, which telling us
the system behaves like a slightly strained pristine graphene.

\begin{figure}[!ht]\centering
    \subfloat[]{
        \includegraphics[scale=0.08]{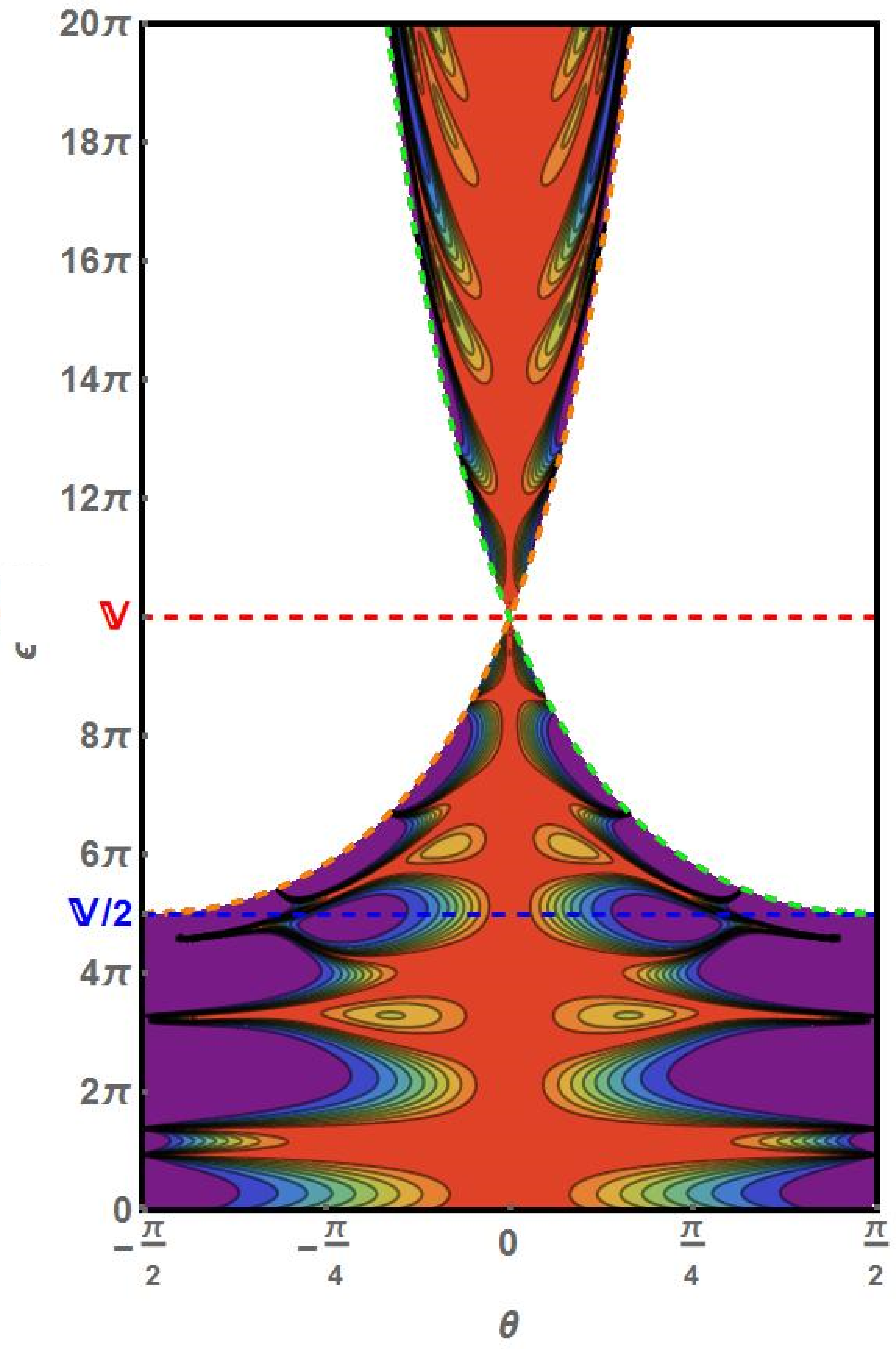}
        \label{fig06:SubFigA}\ \ \ \
    }\hspace{-0.4cm}
    \subfloat[]{
        \includegraphics[scale=0.08]{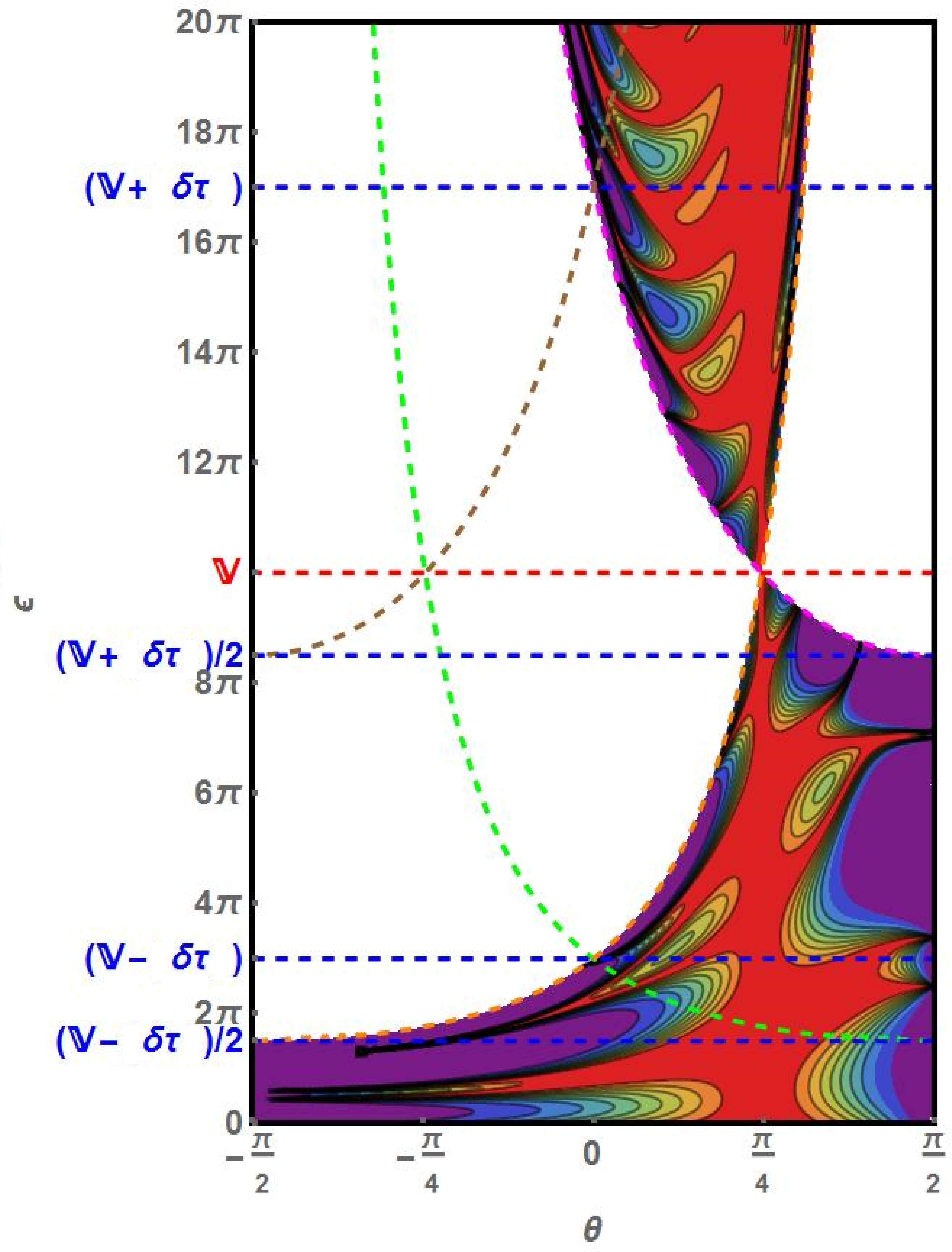}
        \label{fig06:SubFigB} \ \ \ \
    }\hspace{-0.4cm}
    \subfloat[]{
        \includegraphics[scale=0.08]{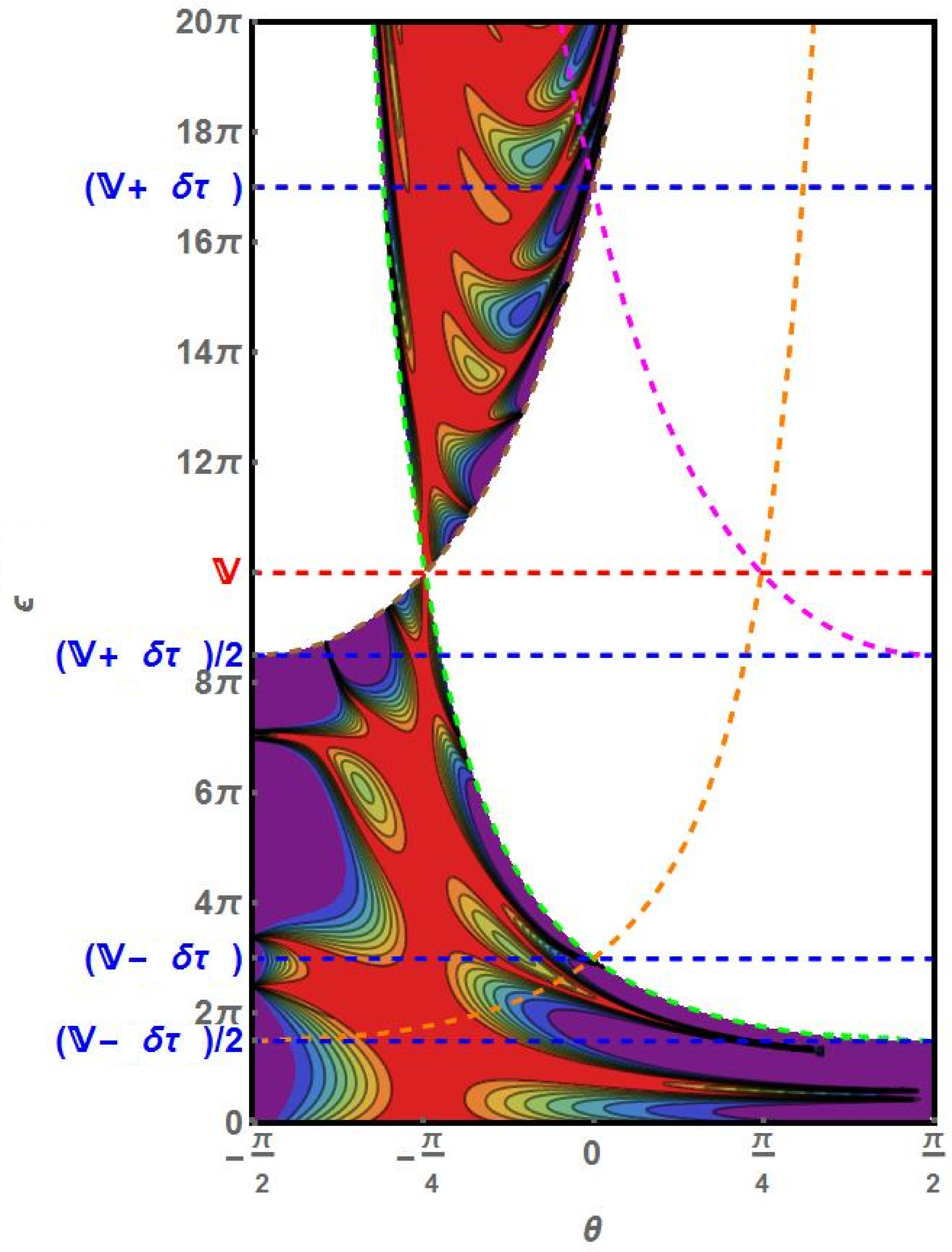}
        \includegraphics[scale=0.15]{0000}
        \label{fig06:SubFigC}
    }\\
    \vspace{-0.4cm}
    \subfloat[]{
        \includegraphics[scale=0.08]{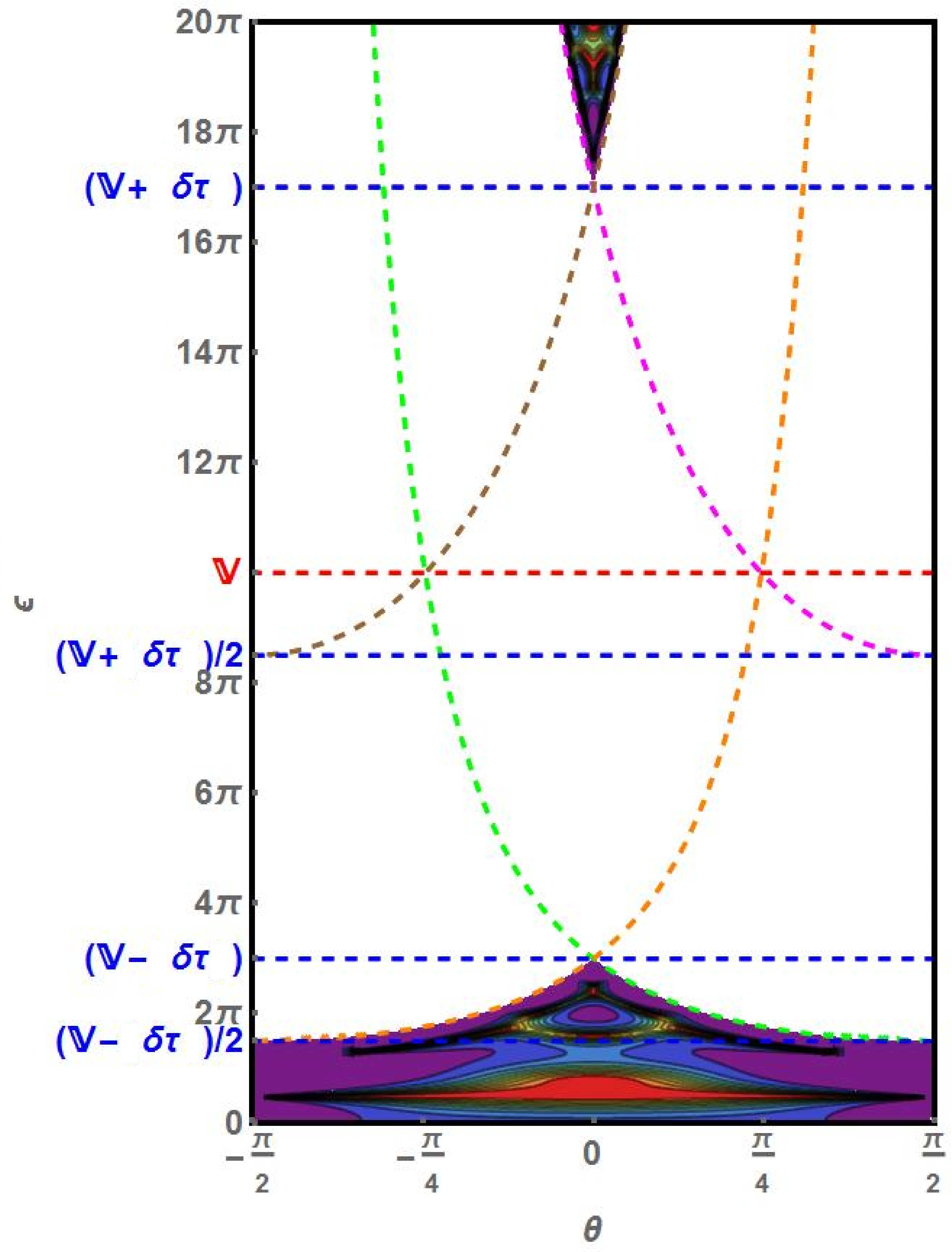}
        \label{fig06:SubFigD} \ \ \ \
    }\hspace{-0.4cm}
    \subfloat[]{
        \includegraphics[scale=0.08]{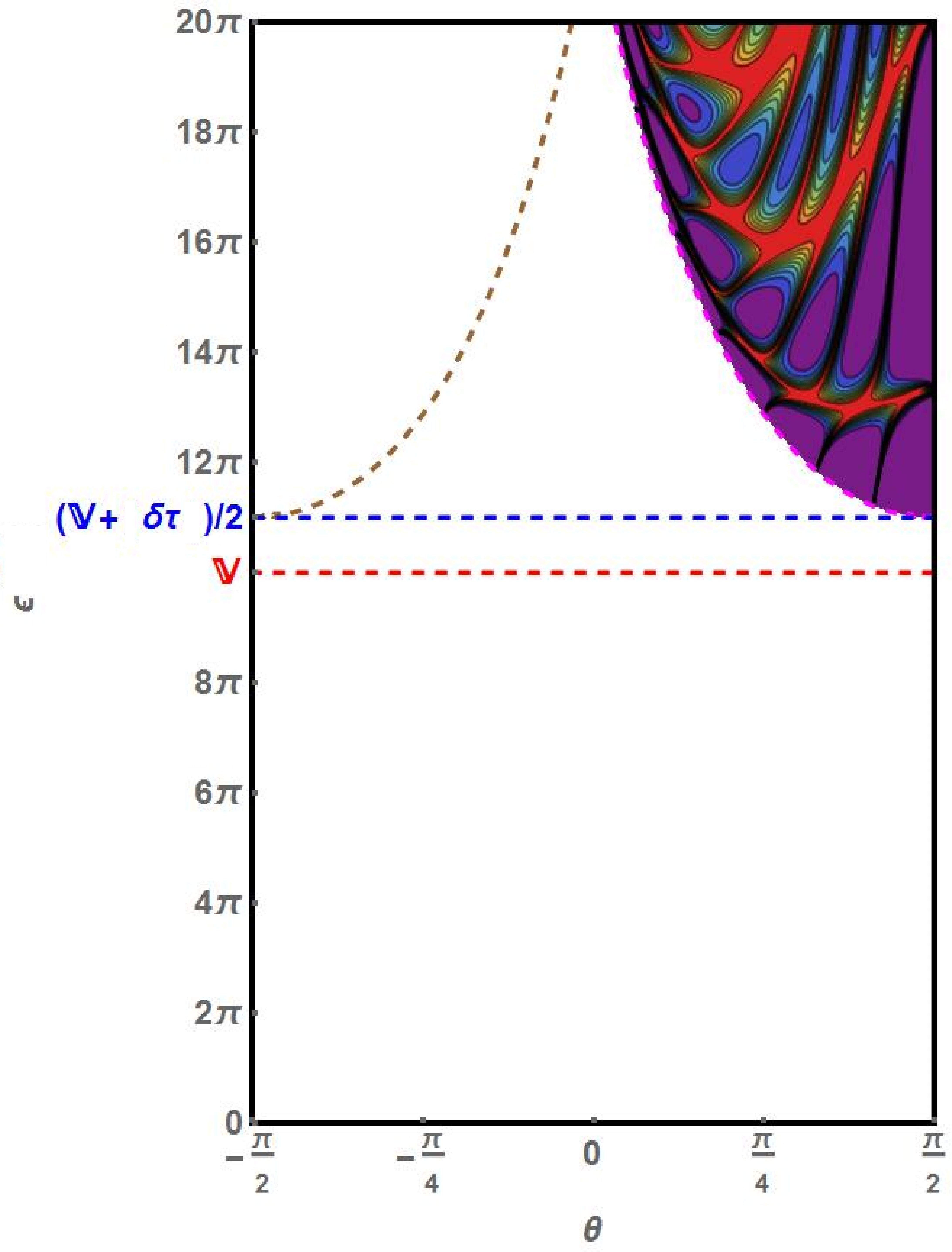}
        \label{fig06:SubFigE} \ \ \ \
    }\hspace{-0.4cm}
        \subfloat[]{
        \includegraphics[scale=0.08]{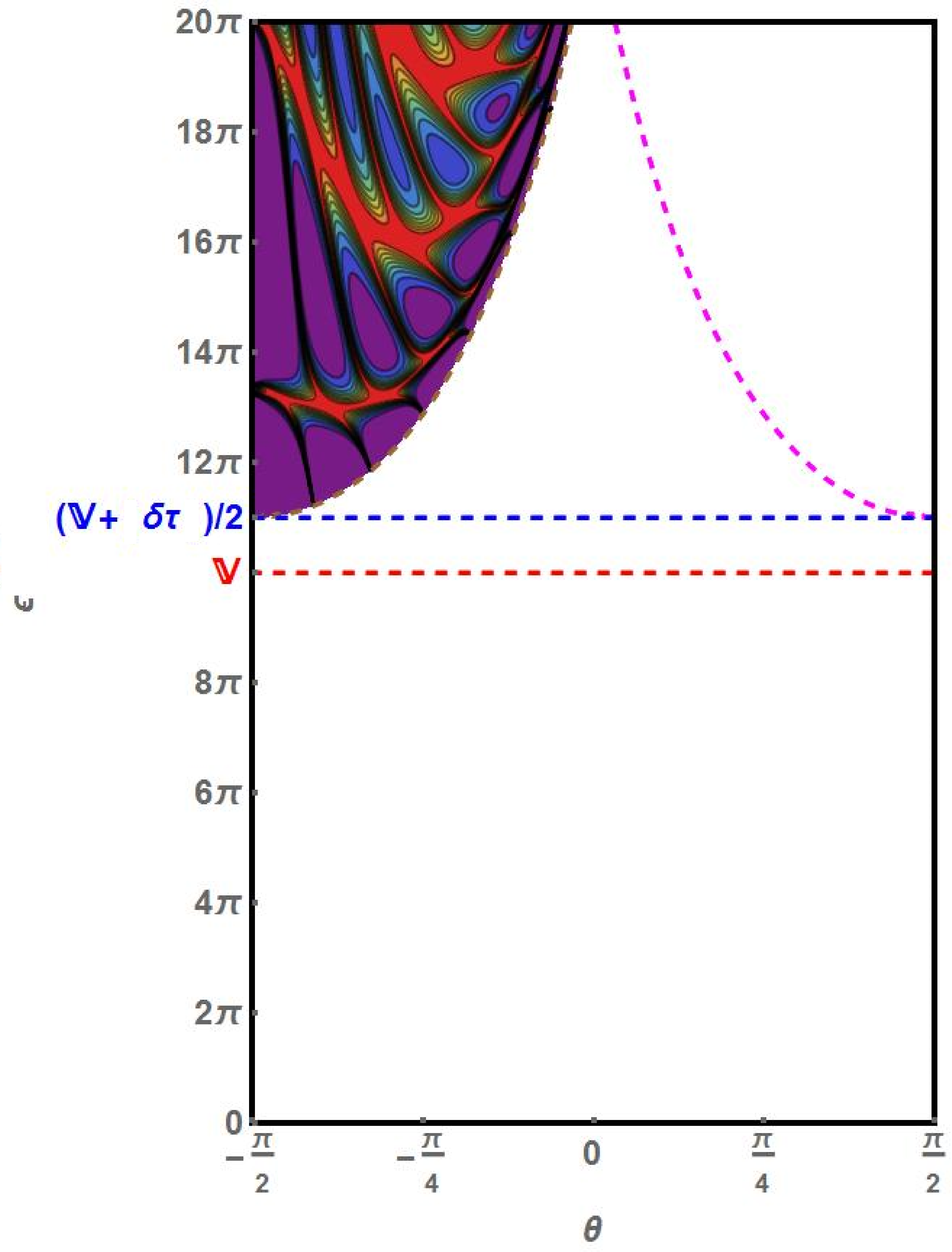}
        \includegraphics[scale=0.15]{0000}
        \label{fig06:SubFigF}
    }
    \caption{
                (color online) Density plot of transmission probability $T$ versus incident
                energy $\varepsilon$ and incident angle $\theta$ with
                $\mathbb{V}_2=\mathbb{V}_4=\mathbb{V}=10 \pi$, 
                $\varrho=300$, $d=3\varrho$, $\mathbb{V}=10\pi$ (red dashed line),
                $\left( \frac{\mathbb{V}\pm \delta \tau}{2},
                 \mathbb{V}\pm \delta \tau\right)$
                (blue dashed lines),
                ${\varepsilon^{+}_{-}=\frac{\mathbb{V}+\delta \tau}{1-\sin\theta}}$
                (brown dashed line), ${\varepsilon^{+}_{+}=\frac{\mathbb{V}+\delta \tau}{1+\sin\theta}}$ 
                (magenta dashed line),
                ${\varepsilon^{-}_{-}=\frac{\mathbb{V}-\delta \tau}{1-\sin\theta}}$ 
                (orange dashed line), ${\varepsilon^{-}_{+}=\frac{\mathbb{V}-\delta \tau}{1+\sin\theta}}$
                (green dashed line).  \protect \subref{fig06:SubFigA}:
                $\delta \tau_2 =\delta \tau_4=\delta \tau=0$. \protect \subref{fig06:SubFigB}:
                $\delta \tau_2 =\delta \tau_4=\delta \tau=7\pi$. \protect \subref{fig06:SubFigC}:
                $\delta \tau_2 =\delta \tau_4=\delta \tau=-7\pi$. \protect \subref{fig06:SubFigD}:
                $\delta \tau_2 =-\delta \tau_4=\delta \tau=\pm7\pi$. \protect \subref{fig06:SubFigE}:
                $\delta \tau_2 =\delta \tau_4=\delta \tau=12\pi$ and ${\delta \tau}> \mathbb{V}$.
                \protect \subref{fig06:SubFigF}: $\delta \tau_2 =\delta \tau_4=\delta \tau=-12\pi$ and ${\delta \tau}> \mathbb{V}$.
        }
        \label{Fig06}
\end{figure}
Beside the strained graphene, a symmetrical potential profile
$(\mathbb{V}_2,\mathbb{V}_4)=(\mathbb{V},\mathbb{V})$
has been applied to study its additional effect compounded to that of strain. Figure \ref{Fig06} shows
the transmission density plot versus incident energy $\varepsilon$ and incident angle $\theta$.
The allowed transmissions correspond, respectively to the energy
zones:
{$[\varepsilon \geq (\varepsilon^{+}_{-}=\varepsilon^{-}_{-})\wedge \varepsilon \geq
(\varepsilon^{+}_{-}=\varepsilon^{-}_{+})] \vee [\varepsilon \leq (\varepsilon^{+}_{-}=\varepsilon^{-}_{-})\wedge
\varepsilon \leq (\varepsilon^{+}_{-}=\varepsilon^{-}_{+})]$}  Figure \ref{Fig06}\subref{fig06:SubFigA},
{$[\varepsilon \geq \varepsilon^{-}_{-} \wedge \varepsilon \geq \varepsilon^{+}_{+}] \vee [\varepsilon \leq \varepsilon^{-}_{-}
\wedge \varepsilon \leq\varepsilon^{+}_{+}]$ } Figure  \ref{Fig06}\subref{fig06:SubFigB}, {$[\varepsilon \geq \varepsilon^{-}_{+}
\wedge \varepsilon \geq \varepsilon^{+}_{-}] \vee [\varepsilon \leq \varepsilon^{-}_{+}\wedge \varepsilon \leq \varepsilon^{+}_{-}]$}
Figure \ref{Fig06}\subref{fig06:SubFigC}, {$[\varepsilon \geq \varepsilon^{+}_{+} \wedge \varepsilon \geq \varepsilon^{+}_{-}]
\vee [\varepsilon \leq \varepsilon^{-}_{+}\wedge \varepsilon \leq \varepsilon^{-}_{-}]$ } Figure \ref{Fig06}\subref{fig06:SubFigD},
{$[\varepsilon \geq \varepsilon^{+}_{+} ]$ } Figure \ref{Fig06}\subref{fig06:SubFigE}, and {$[\varepsilon \geq \varepsilon^{+}_{+} ]$}
Figure \ref{Fig06}\subref{fig06:SubFigF}. The forbidden transmission zones (white color)
correspond respectively to the complementary allowed transmission zones  in each systems as illustrated
in Figure \ref{Fig06}, with ${\varepsilon^{\pm}_{+}=\frac{\mathbb{V}\pm \delta \tau}{1+\sin\theta}}$
and ${\varepsilon^{\pm}_{-}=\frac{\mathbb{V}\pm \delta \tau}{1-\sin\theta}}$.
 Figure \ref{Fig06}\subref{fig06:SubFigA} shows the transmission probability
 $T(\varepsilon,\theta)$ for a strainless double barrier with potential
 ($\mathbb{V}_{2}=\mathbb{V}$), the energy spectrum is subdivided
 into three energy domains. The first domain is {$0\leq\varepsilon\leq\frac{\mathbb{V}}{2}$}
 where the transmission is allowed for all incidence angles $-\frac{\pi}{2}\leq \theta\leq \frac{\pi}{2}$,
 the double barrier  behaves as a more refractive medium than pristine graphene.
 The second domain is {$\frac{\mathbb{V}}{2}\leq\varepsilon\leq\mathbb{V}$}, which shows that at each energy
 there are two critical angles symmetrical with respect to the normal incidence,
 they come closer   when the energy tends to $\mathbb{V}$. The third domain
 corresponds to { $\varepsilon \geq \mathbb{V}$}, when the energy increases
 the critical angles remain always symmetrical but they move away  parabolically.
 In the two last domains, the double barrier behaves like a less refractive medium than pristine graphene.
In vicinity of normal incidence, we have a total transmission (red color) whatever
the propagating energy of Dirac fermions (Klein Paradox). The transmission probability
is symmetrical with respect to normal incidence angle $\theta=0$, it vanishes when approaching the limit angles
to give transmission gaps (purple color) intercalated between resonance peaks.
The transmission, in Figure \ref{Fig06}\subref{fig06:SubFigA}, has
the same form  as that obtained by  the transmission through a potential
barrier in monolayer graphene studied in \cite{Chinese}.
 In Figures (\ref{Fig06}\subref{fig06:SubFigB}, \ref{Fig06}\subref{fig06:SubFigC}),
 we illustrate the double barrier, which becomes strained by adding
 $(\delta \tau_2 =\delta \tau_4=\delta \tau=7\pi,\delta \tau<\mathbb{V})$
 and $(\delta \tau_2 =\delta \tau_4=\delta \tau=-7\pi,\delta \tau<\mathbb{V})$,
 respectively. With respect to normal incidence, the two transmissions in these
 last two Figures are symmetrical to each other, the transmission lose its symmetry
 (as compared to Figure \ref{Fig06}\subref{fig06:SubFigA}) with respect to the normal  incidence.
 The transmission has an additional domain compared to that of strainless double barrier,
 which just appeared between the first and the second zones. In this domain,
 the transmission in \ref{Fig06}\subref{fig06:SubFigB} (\ref{Fig06}\subref{fig06:SubFigC})
 is allowed in the positive (negative) incidence $\theta>0$ ($\theta<0$)
 and limited by critical angles in the negative (positive) incidence $\theta<0$ ($\theta>0$).
 In Figures (\ref{Fig06}\subref{fig06:SubFigE}, \ref{Fig06}\subref{fig06:SubFigF})
 plotted 
 for
 $(\delta \tau_2 =\delta \tau_4=\delta \tau=12\pi,\delta \tau>\mathbb{V})$ and
 $(\delta \tau_2 =\delta \tau_4=\delta \tau=-12\pi,\delta \tau>\mathbb{V})$,
 respectively, there remains only one transmission domain for energies
 $\varepsilon>\frac{\mathbb{V}+\delta \tau}{2}$. The system behaves
 as a strained double  barrier, with no potential profile, with a like
 compression strain $\delta \tau'=\mathbb{V}+\delta \tau$.
Figure \ref{Fig06}\subref{fig06:SubFigD} for a double
barrier composed by compression and traction
($\delta \tau_2 =-\delta \tau_4=\delta \tau=\pm7\pi$) with
a predicted potential $(\mathbb{V}_2,\mathbb{V}_4)=(\mathbb{V},\mathbb{V})$,
shows an energy gap between $\varepsilon=\mathbb{V}+\delta \tau$
and $\varepsilon=\mathbb{V}-\delta \tau$ of width $2\delta \tau$.
For energy {$0\leq\varepsilon\leq\frac{\mathbb{V}-\delta \tau}{2}$},
the double barrier behaves like a more refractive medium than pristine graphene.
On the other hand, for the energy { $\frac{\mathbb{V}-\delta \tau}{2}\leq\varepsilon\leq\mathbb{V}-\delta \tau$},
it behaves like a less refractive medium and the critical angles are symmetric with respect
to $\theta=0$. The critical angles for  energy {$\varepsilon\geq \mathbb{V}+\delta \tau$}
are also symmetrical and increase as a function of energy in a parabolic way.

\begin{figure}[!ht]\centering
    \subfloat[]{
        \includegraphics[scale=0.33]{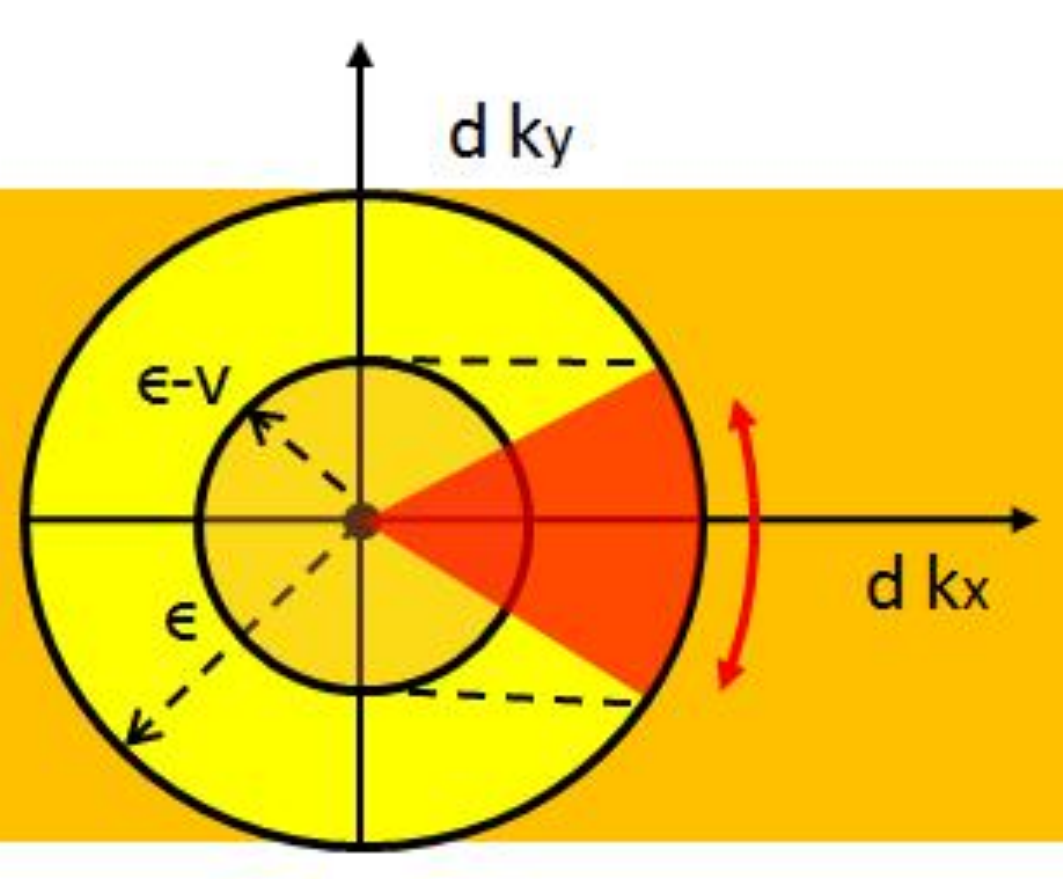}
        \label{fig07:SubFigA}\ \ \ \
    }\hspace{-0.2cm}
    \subfloat[]{
        \includegraphics[scale=0.33]{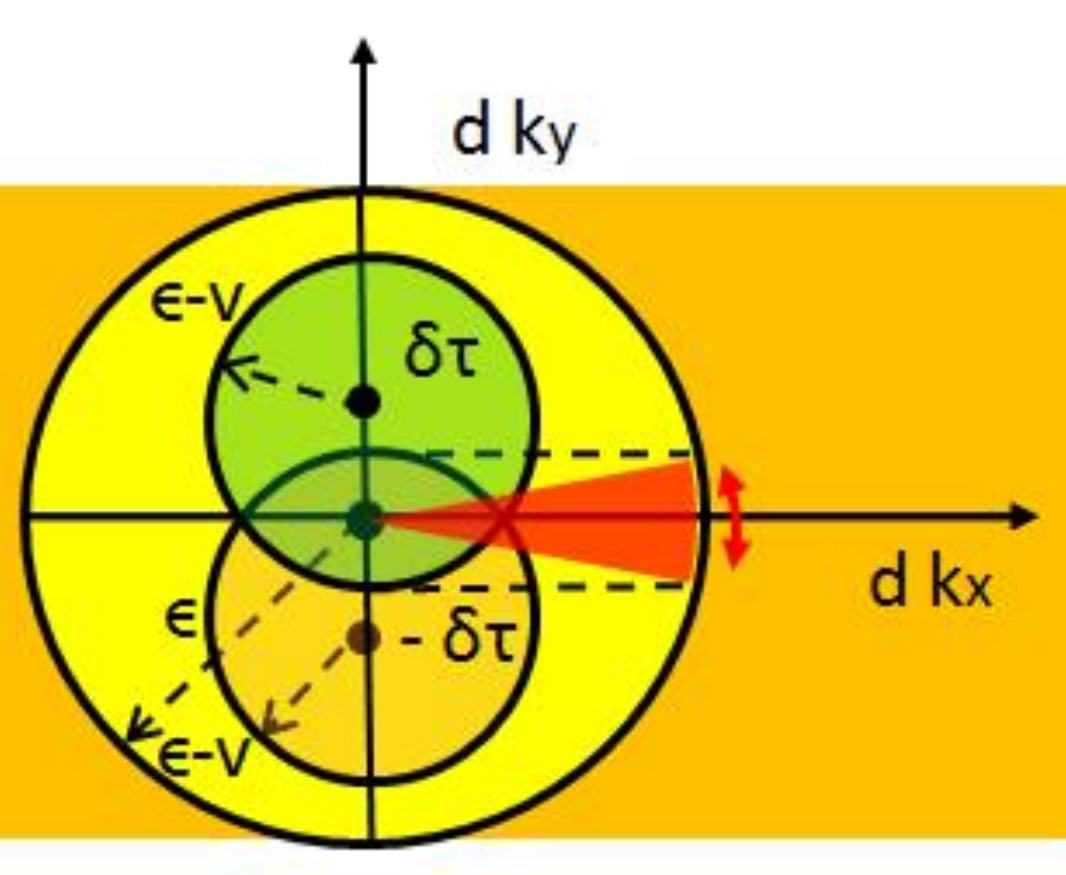}
        \label{fig07:SubFigB} \ \ \ \
    }\hspace{-0.2cm}
        \subfloat[]{
        \includegraphics[scale=0.33]{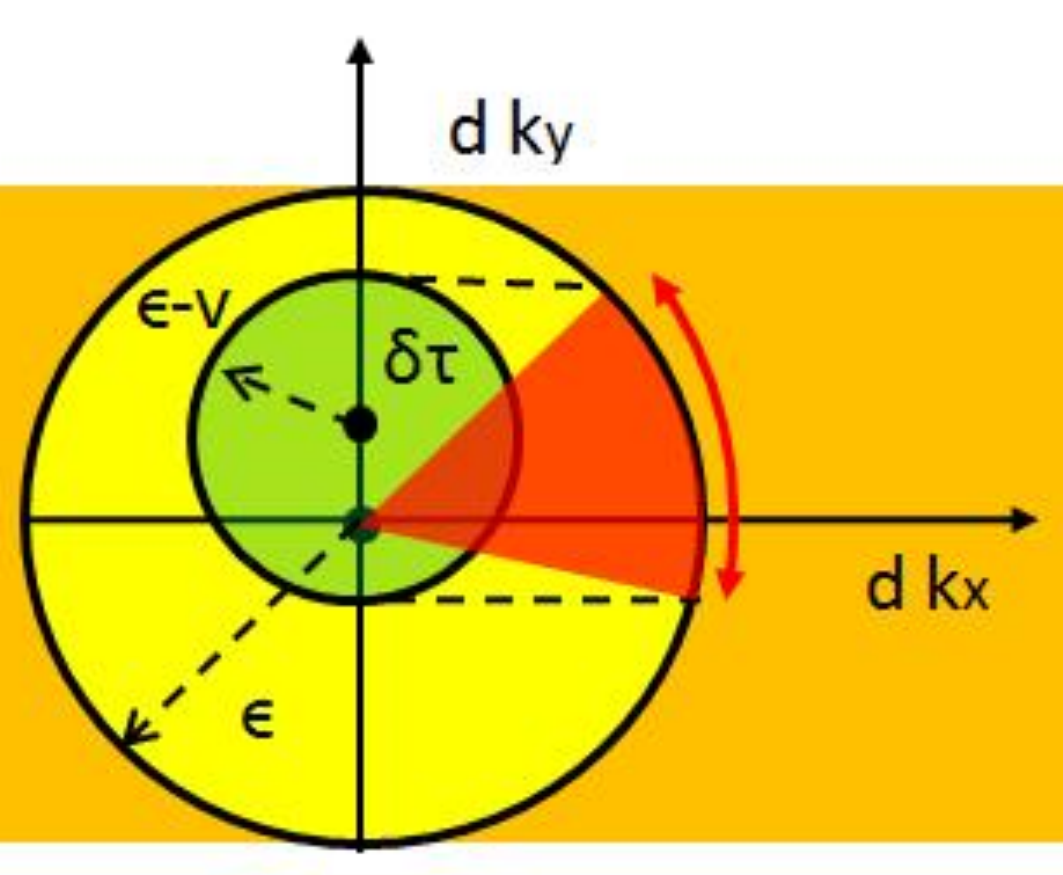}
        \label{fig07:SubFigC}\ \ \ \
    }
    \hspace{-0.2cm}
        \subfloat[]{
        \includegraphics[scale=0.33]{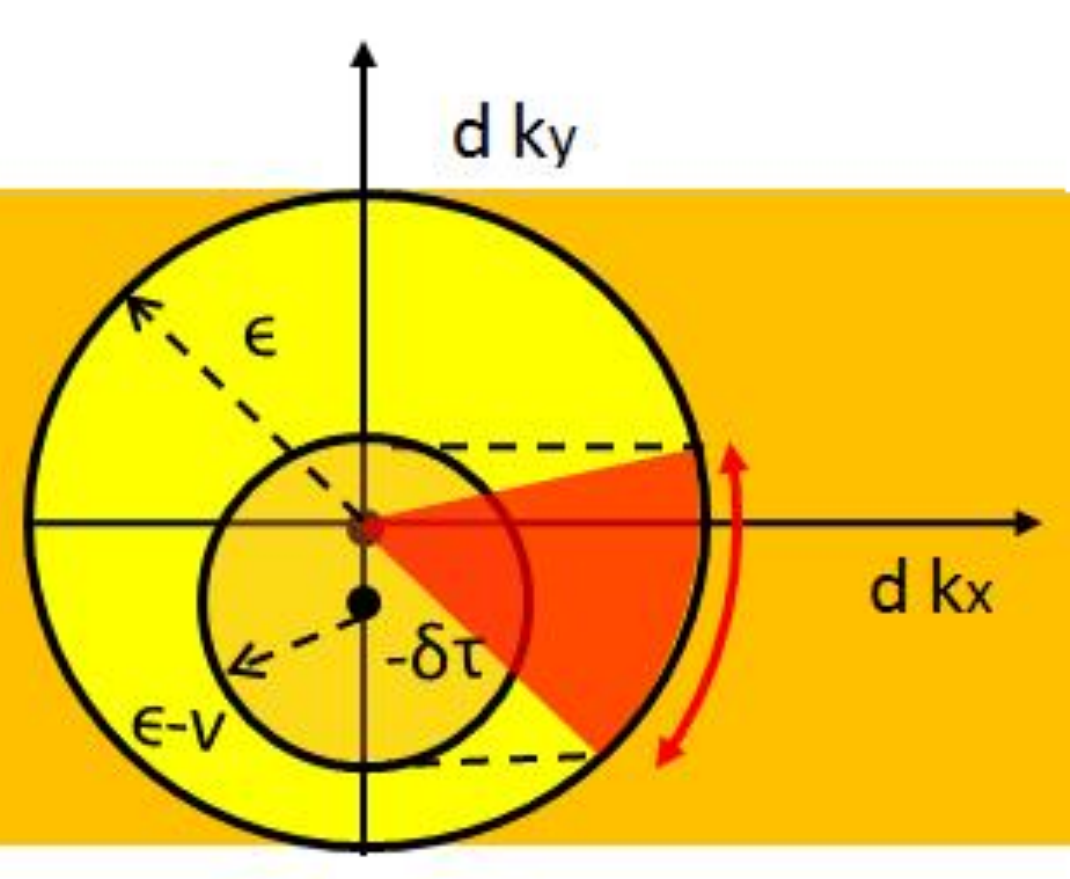}
        \label{fig07:SubFigD}
    }
    \caption{
                (color online) Fermi surfaces affected by strain and potential $\mathbb{V}$ in double barrier.
                \protect \subref{fig07:SubFigA}:
                $\delta \tau_2 =\delta \tau_4=0$. \protect \subref{fig07:SubFigB}:
                $\delta \tau_2 =-\delta \tau_4=\pm\delta \tau$.
                \protect \subref{fig07:SubFigC}: $\delta \tau_2 =\delta \tau_4=\delta \tau$.
        }
        \label{Fig07}
\end{figure}

By adding the potential to strained graphene (Figure \ref{Fig07}), the Fermi surfaces of
each region $\textbf{\textcircled{2}}$ and $\textbf{\textcircled{4}}$ change their radii
and their equations become respectively $(\varepsilon-\mathbb{V})^{2}=d^{2}k_2^{2}+(d k_y\pm\delta\tau)^{2}$ and
$(\varepsilon-\mathbb{V})^{2}=d^{2}k_4^{2}+(d k_y\pm\delta\tau)^{2}$.
Since the radius becomes variable with the potential $\mathbb{V}$, Figure \ref{Fig07} generalizes
Figure \ref{Fig02}. The collimation realized by filtering of certain angles of incidence \cite{8}
for  different configurations of the double strained barrier, makes it possible to have
the contours delimiting the allowed and forbidden transmission zones (red zones)
in Figure  \ref{Fig07}.
Figure \ref{Fig07}\subref{fig07:SubFigA}
illustrates the behavior of Fermi surfaces for a double barrier without strain,
the allowed angles  are symmetric with respect to the normal incidence angle and correspond to
those of the Figure  \ref{Fig06}\subref{fig06:SubFigA}. 
Figure \ref{Fig07}\subref{fig07:SubFigB} illustrates the behavior of the Fermi surfaces
for a double barrier with two strains of different nature (one of contraction and the other of compression),
the allowed angles  have also a  symmetry with respect to the normal incidence angle
and correspond to those of Figure \ref{Fig06}\subref{fig06:SubFigD}.
Finally Figure \ref{Fig07}\subref{fig06:SubFigC} (\ref{Fig07}\subref{fig06:SubFigD})
shows the Fermi surfaces behavior  of a barrier with two strains of the same nature compression (traction).
The displacement of the origin of Fermi surface  corresponding   to the strained zone, omitted the symmetry
with respect to the normal incidence angle. But the permitted angles of Figure \ref{Fig07}\subref{fig06:SubFigC}
(compression strain) are symmetrical compared to the allowed angles of Figure \ref{Fig07}\subref{fig06:SubFigD}
(traction strain) with respect to the normal angle of incidence.

\begin{figure}[!ht]\centering
        \subfloat[]{
        \includegraphics[scale=0.08]{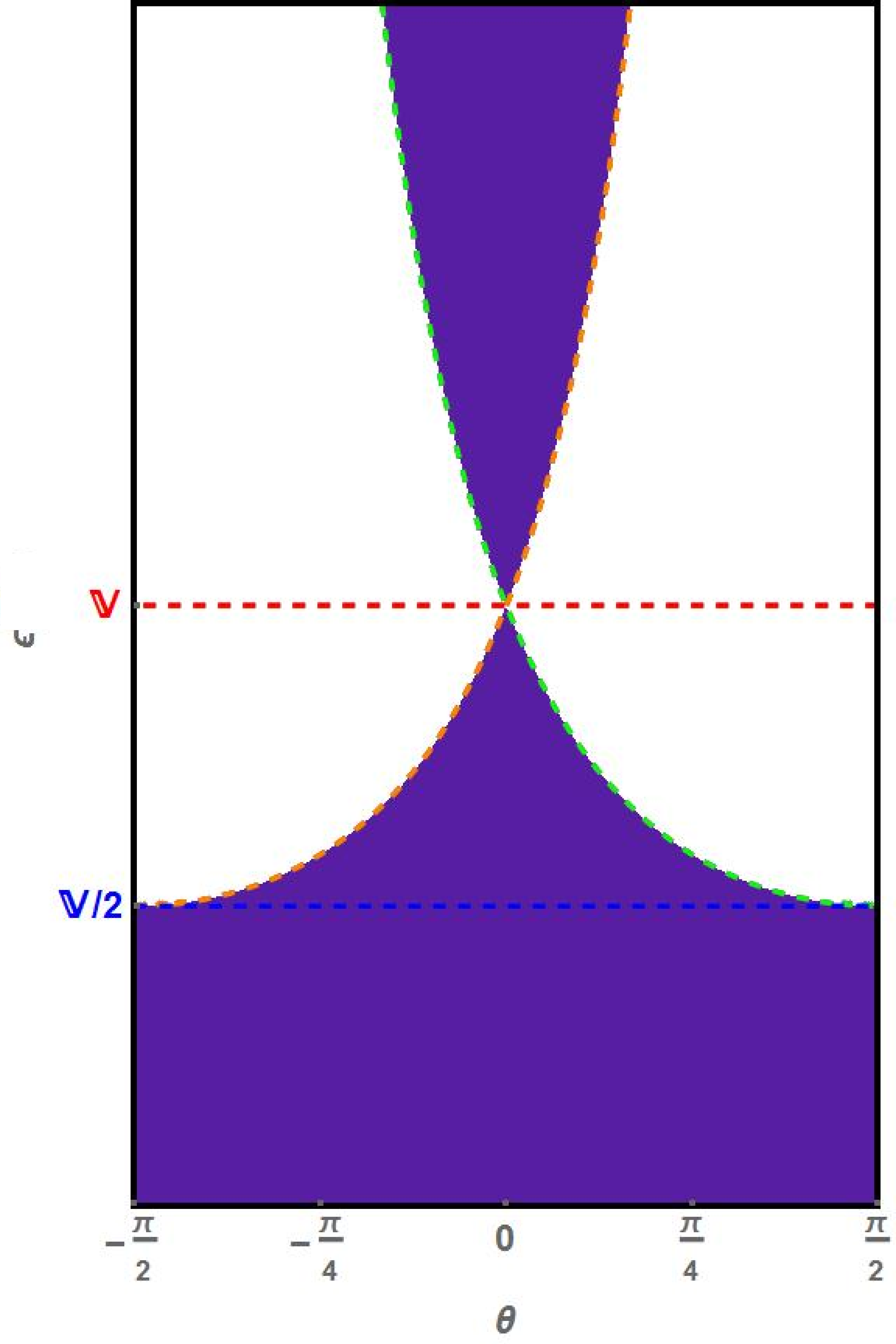}
        \label{fig09:SubFigA}\ \ \ \
    }\hspace{-0.02cm}
        \subfloat[]{
        \includegraphics[scale=0.08]{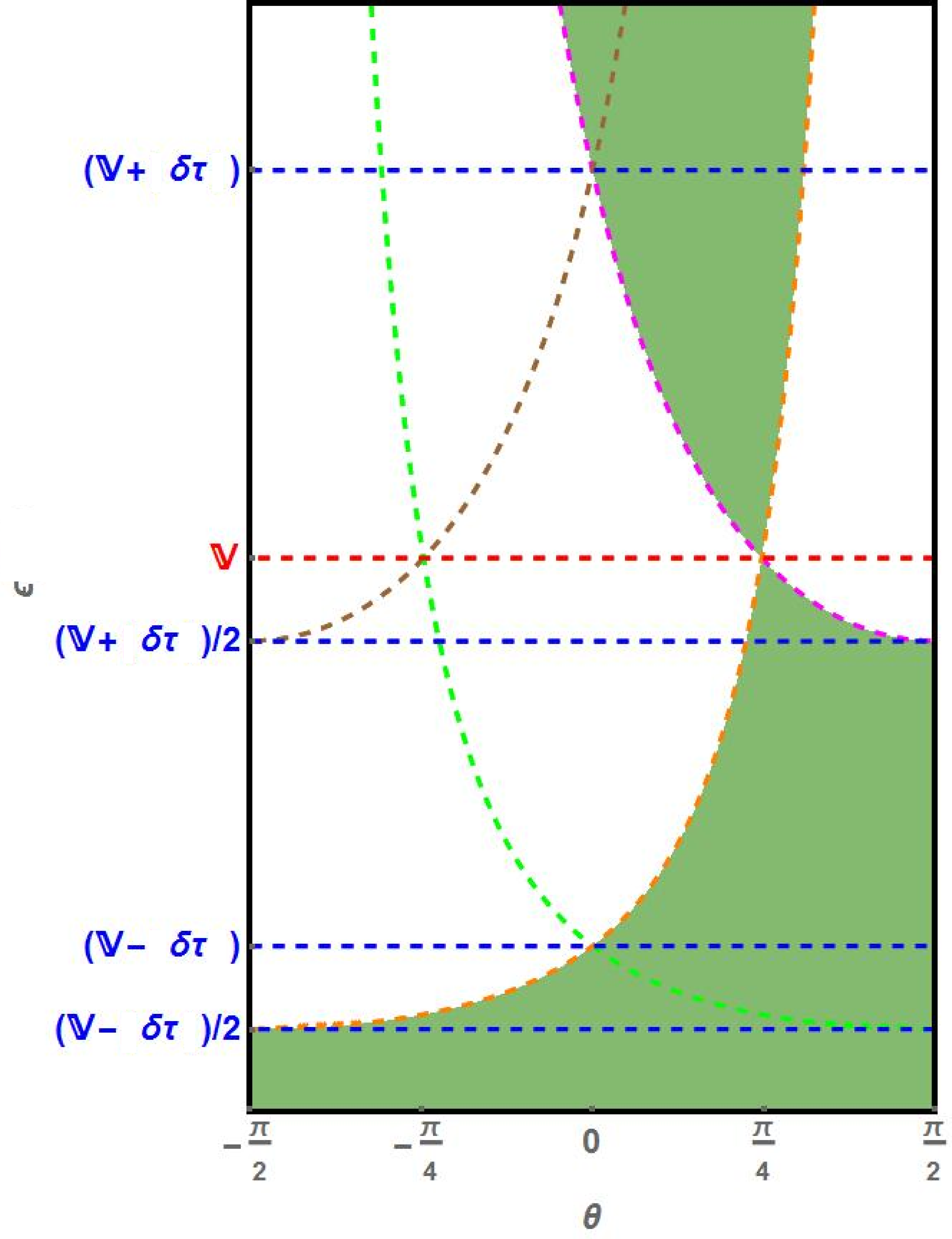}
        \label{fig09:SubFigB}\ \ \ \
    }\hspace{-0.02cm}
        \subfloat[]{
        \includegraphics[scale=0.08]{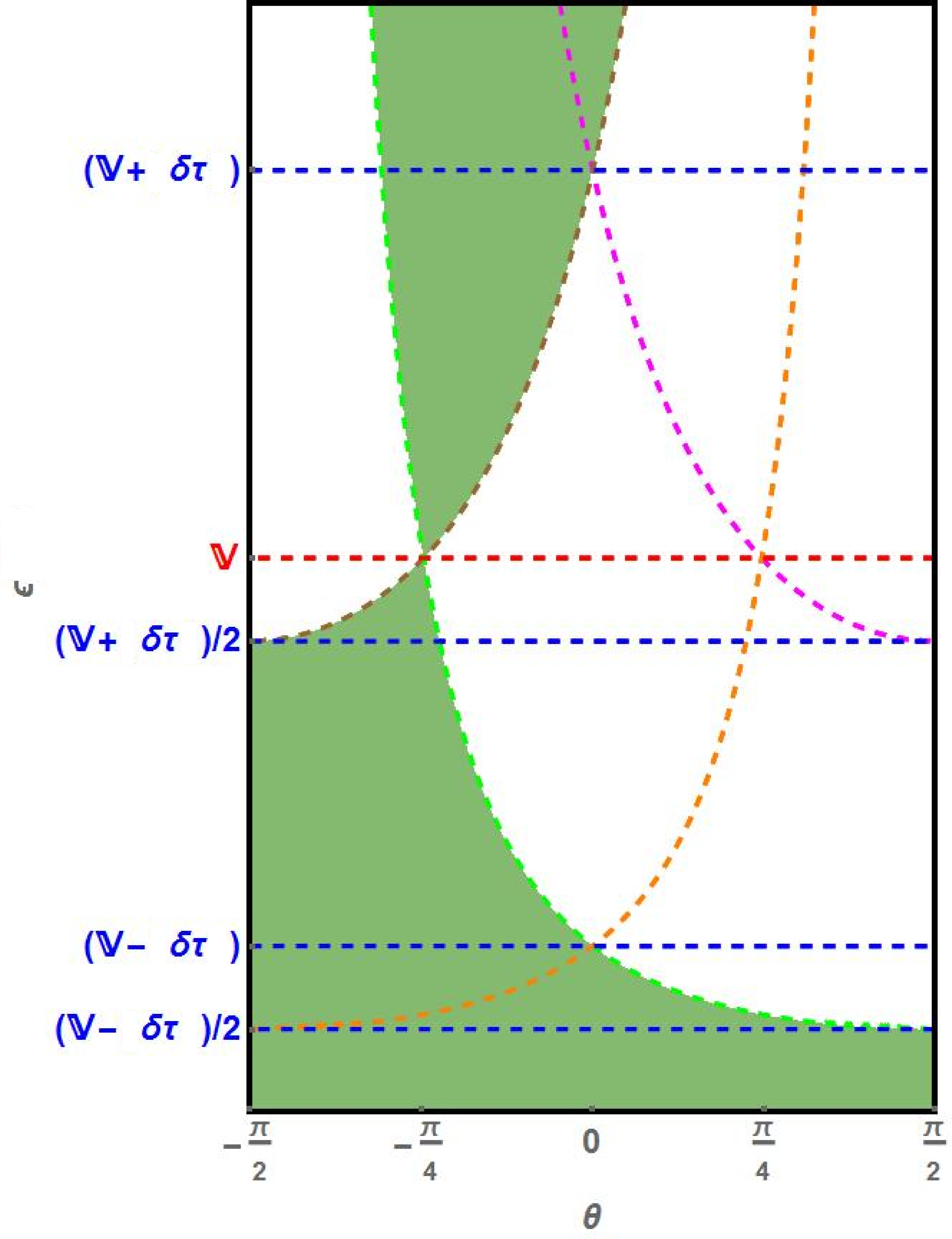}
        \label{fig09:SubFigC}
    }\\
    \vspace{-0.4cm}
        \subfloat[]{
        \includegraphics[scale=0.08]{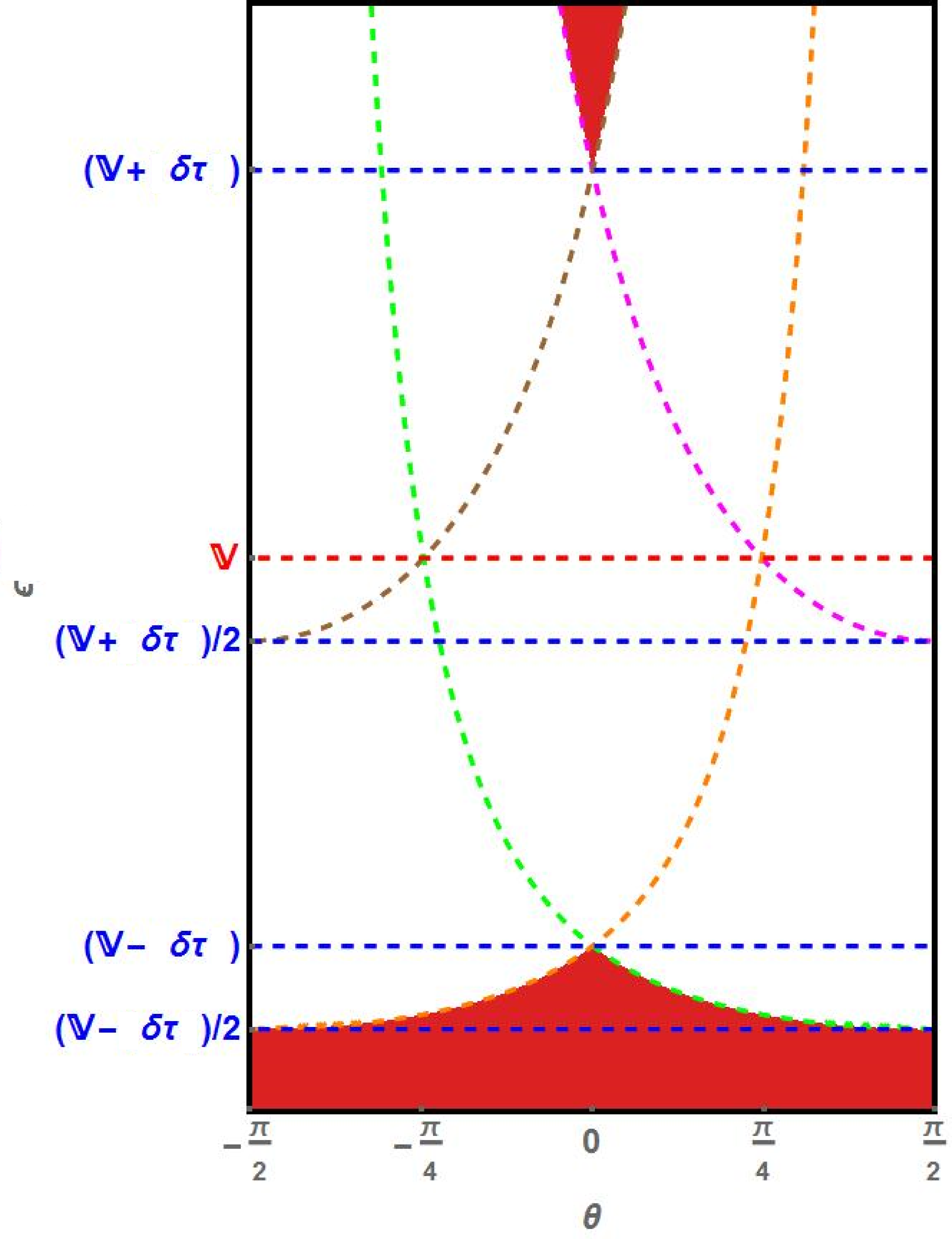}
        \label{fig09:SubFigD}\ \ \ \
    }\hspace{-0.02cm}
        \subfloat[]{
        \includegraphics[scale=0.08]{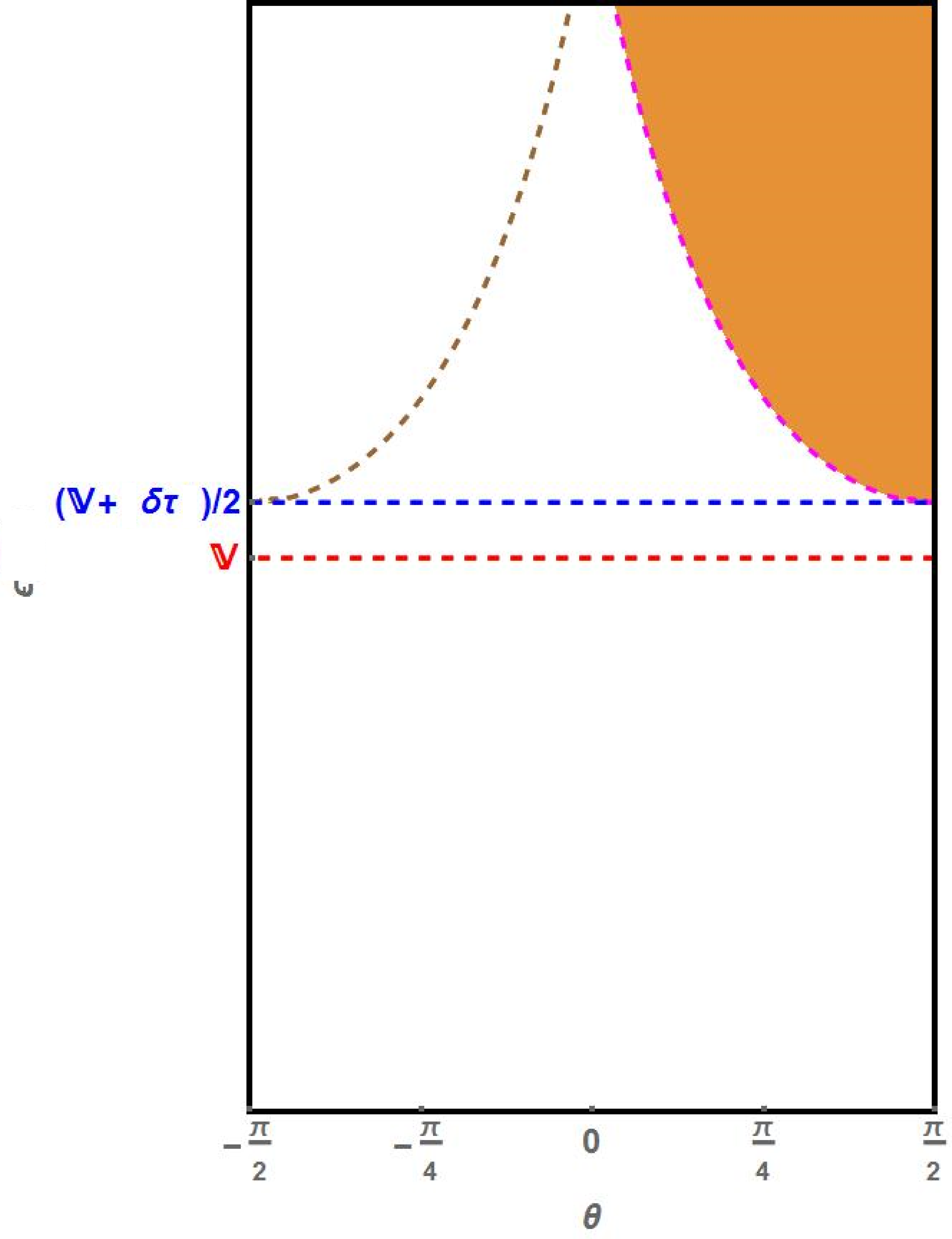}
        \label{fig09:SubFigE}\ \ \ \
    }
    \hspace{-0.02cm}
        \subfloat[]{
        \includegraphics[scale=0.08]{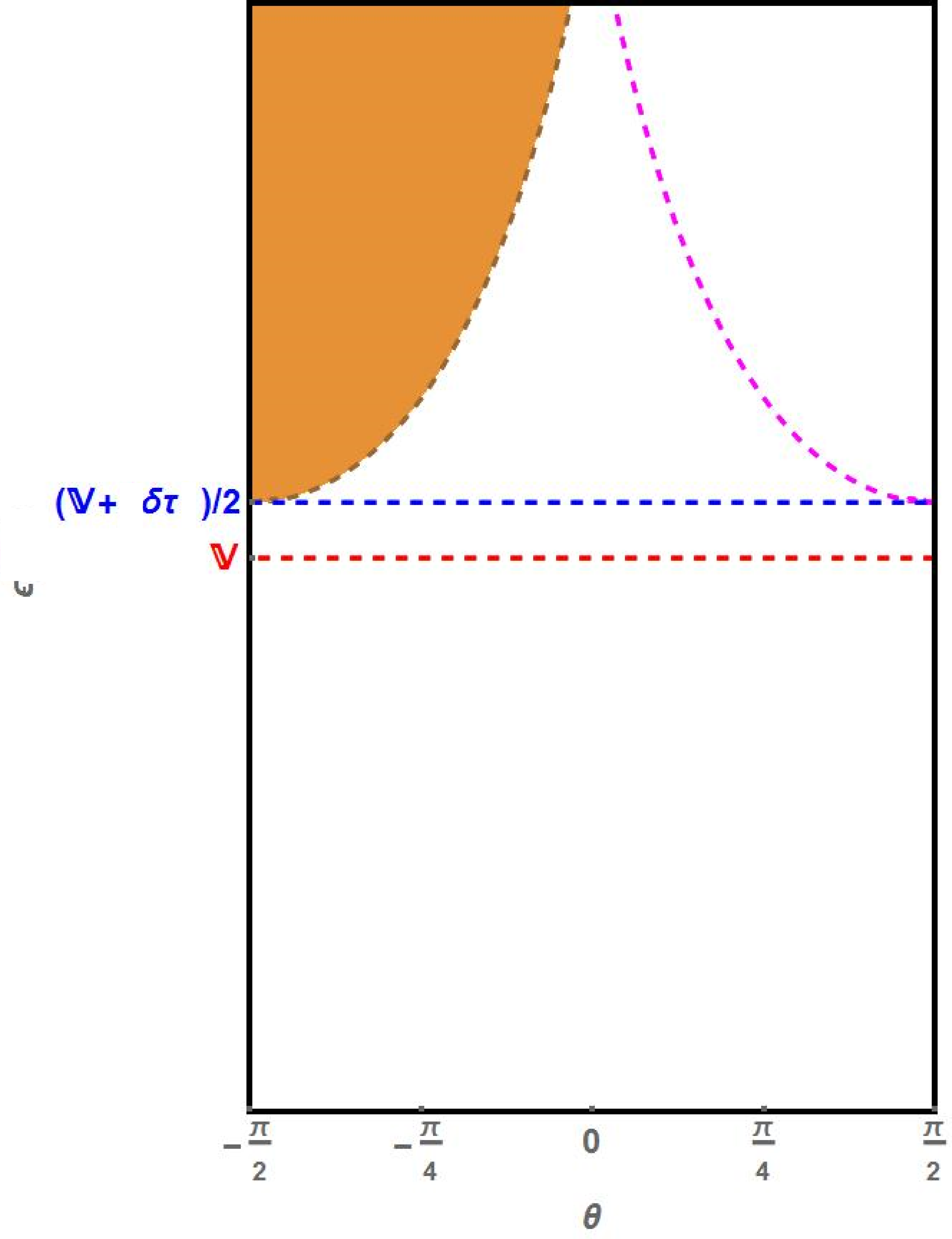}
        \label{fig09:SubFigF}
    }
        \caption{
                (color online) Collimation effect on the energy zones, in strained
                double barrier under the potential profile $(\mathbb{V}_2,\mathbb{V}_4)=(\mathbb{V},\mathbb{V})$,
                the forbidden energy zones are white and the allowed energy zones are colored,
                $\varepsilon^{+}_{+}\lga {\theta^{+}_{+}= \arcsin \left( \frac{-\varepsilon+\mathbb{V}+\delta\tau}{\varepsilon}\right)}$
                (magenta dashed line), $\varepsilon^{+}_{-}\lga {\theta^{+}_{-}= \arcsin \left( \frac{\varepsilon-\mathbb{V}-\delta\tau}{\varepsilon}\right)}$
                (brown dashed line), $\varepsilon^{-}_{-}\lga {\theta^{-}_{-}= \arcsin \left( \frac{-\varepsilon+\mathbb{V}+\delta\tau}{\varepsilon}\right)}$
                (orange dashed line), $\varepsilon^{-}_{+}\lga {\theta^{-}_{+}= \arcsin \left( \frac{-\varepsilon+\mathbb{V}-\delta\tau}{\varepsilon}\right)}$ (green dashed line).
                \protect \subref{fig09:SubFigA}:
                $\delta \tau_2 =\delta \tau_4=\delta \tau=0$. \protect \subref{fig09:SubFigB}:
                $\delta \tau_2 =\delta \tau_4=\delta \tau=7\pi$. \protect \subref{fig09:SubFigC}:
                $\delta \tau_2 =\delta \tau_4=\delta \tau=-7\pi$. \protect \subref{fig09:SubFigD}:
                $\delta \tau_2 =-\delta \tau_4=\delta \tau=\pm7\pi$. \protect \subref{fig09:SubFigE}:
                $\delta \tau_2 =\delta \tau_4=\delta \tau=12\pi$ and ${\delta \tau}> \mathbb{V}$.
                \protect \subref{fig09:SubFigF}: $\delta \tau_2 =\delta \tau_4=\delta \tau=-12\pi$ and $\delta \tau> \mathbb{V}$.
        }
        \label{Fig09}
\end{figure}

A geometric manipulation is used to determine easily
the limiting angles $\theta^{s}_{s'}=
   \arcsin\left( \frac{\mathbb{V}-\varepsilon+s\delta\tau}{s'\varepsilon}\right)$
   corresponding to energy contours
 $\varepsilon^{s}_{s'}= \frac{\mathbb{V}+s\delta\tau}{ 1+ s'\sin\theta}$ 
separating
 different transmission zones in  Figure \ref{Fig07}, with 
  $s=\pm$ and $s'=\pm$. These
angles 
are well illustrated in  Figure \ref{Fig09} under suitable conditions. Then we have easily
obtained
such angles delimiting  different transmission zones
such as $\varepsilon^{+}_{+}\lga  \theta^{+}_{+}$ (magenta dashed line),
$\varepsilon^{+}_{-}\lga \theta^{+}_{-}$
(brown dashed line),
$\varepsilon^{-}_{-}\lga \theta^{-}_{-}$ 
(orange dashed line),
$\varepsilon^{-}_{+}\lga \theta^{-}_{+}$
(green dashed line).

\begin{figure}[!ht]\centering
    \subfloat[]{
        \includegraphics[scale=0.15]{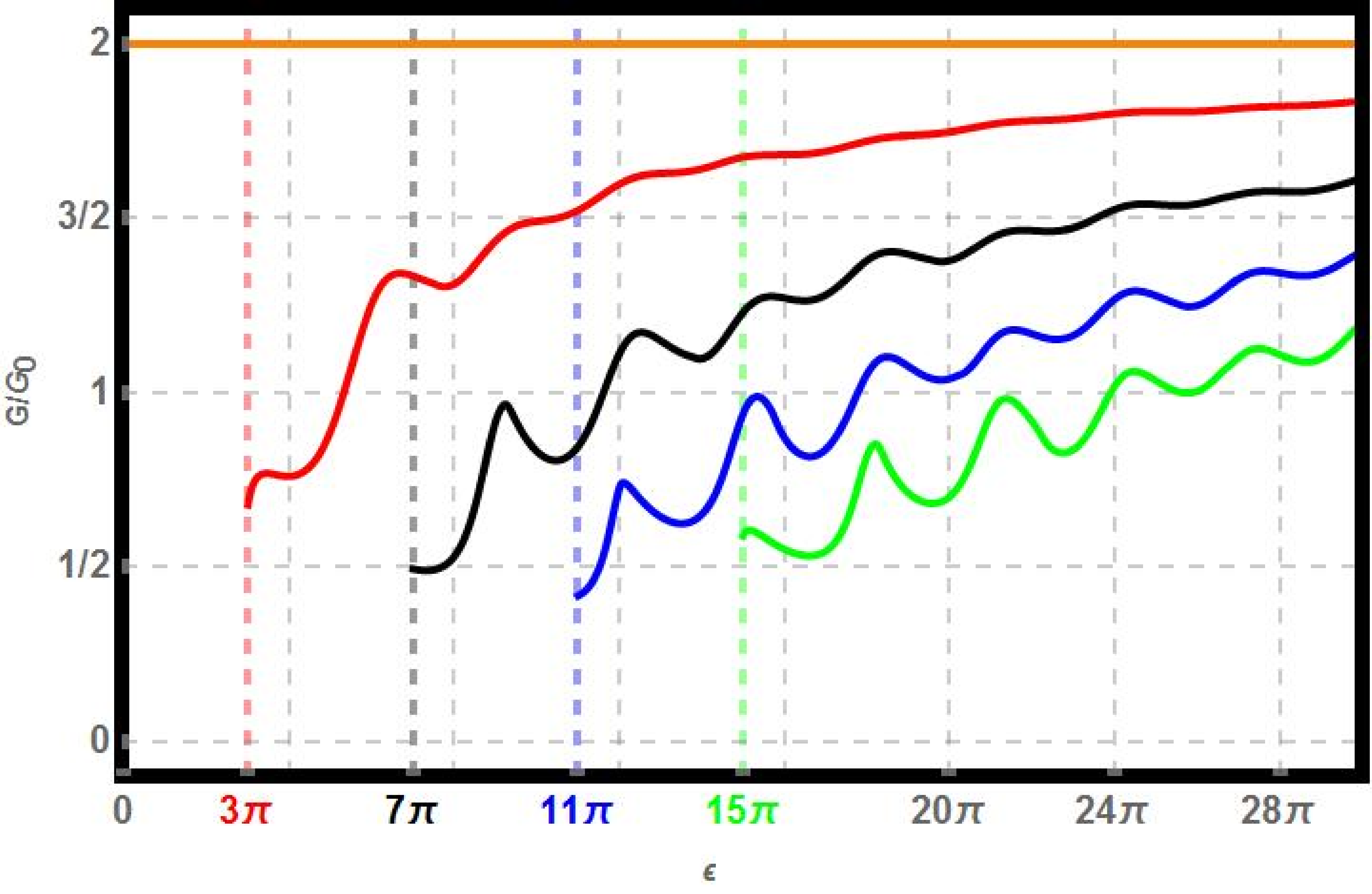}
        \label{fig10:SubFigA}
    }\hspace{-0.02cm}
        \subfloat[]{
        \includegraphics[scale=0.15]{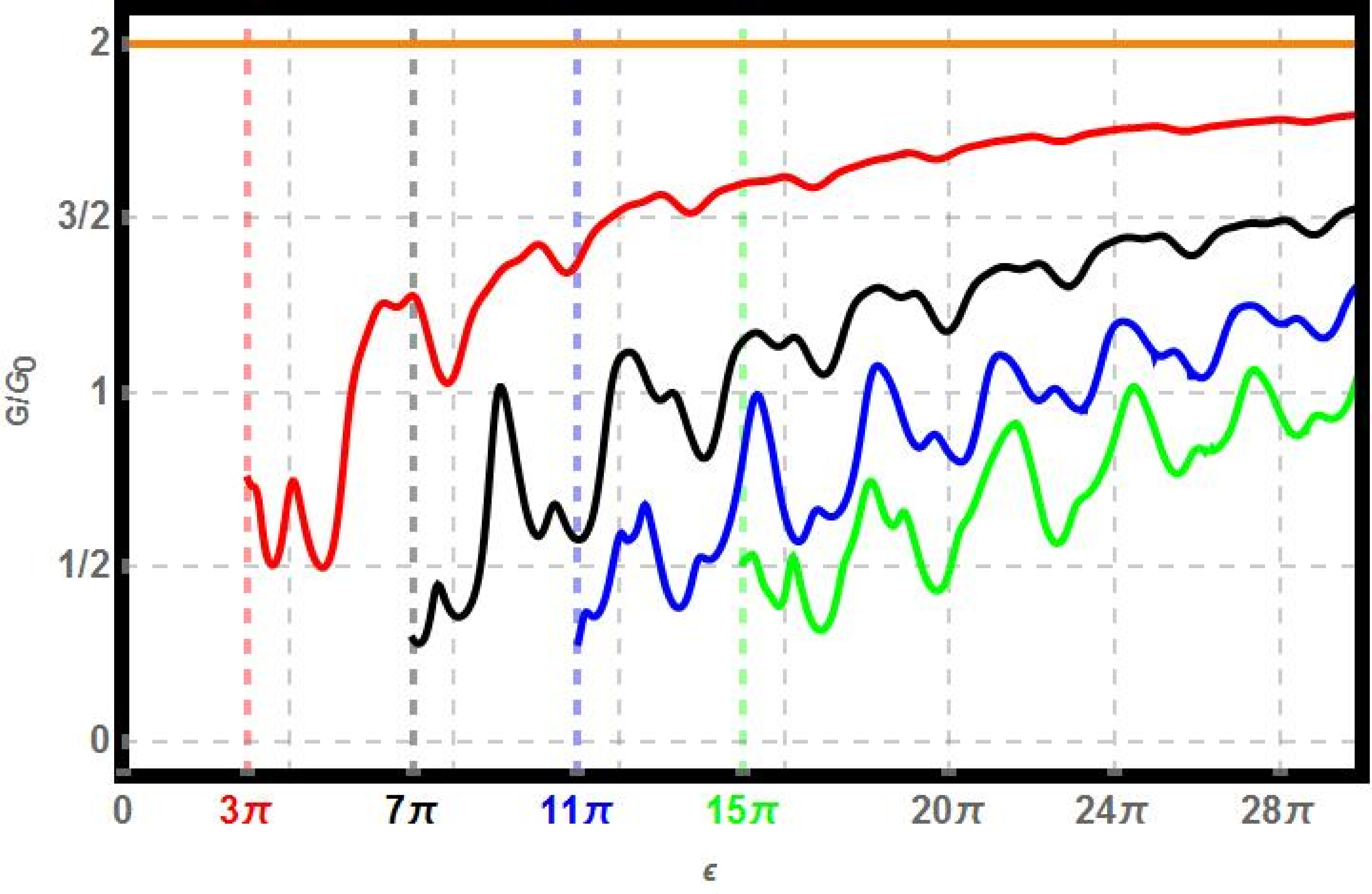}
        \label{fig10:SubFigB}
    }\vspace{-0.4cm}
        \subfloat[]{
        \includegraphics[scale=0.15]{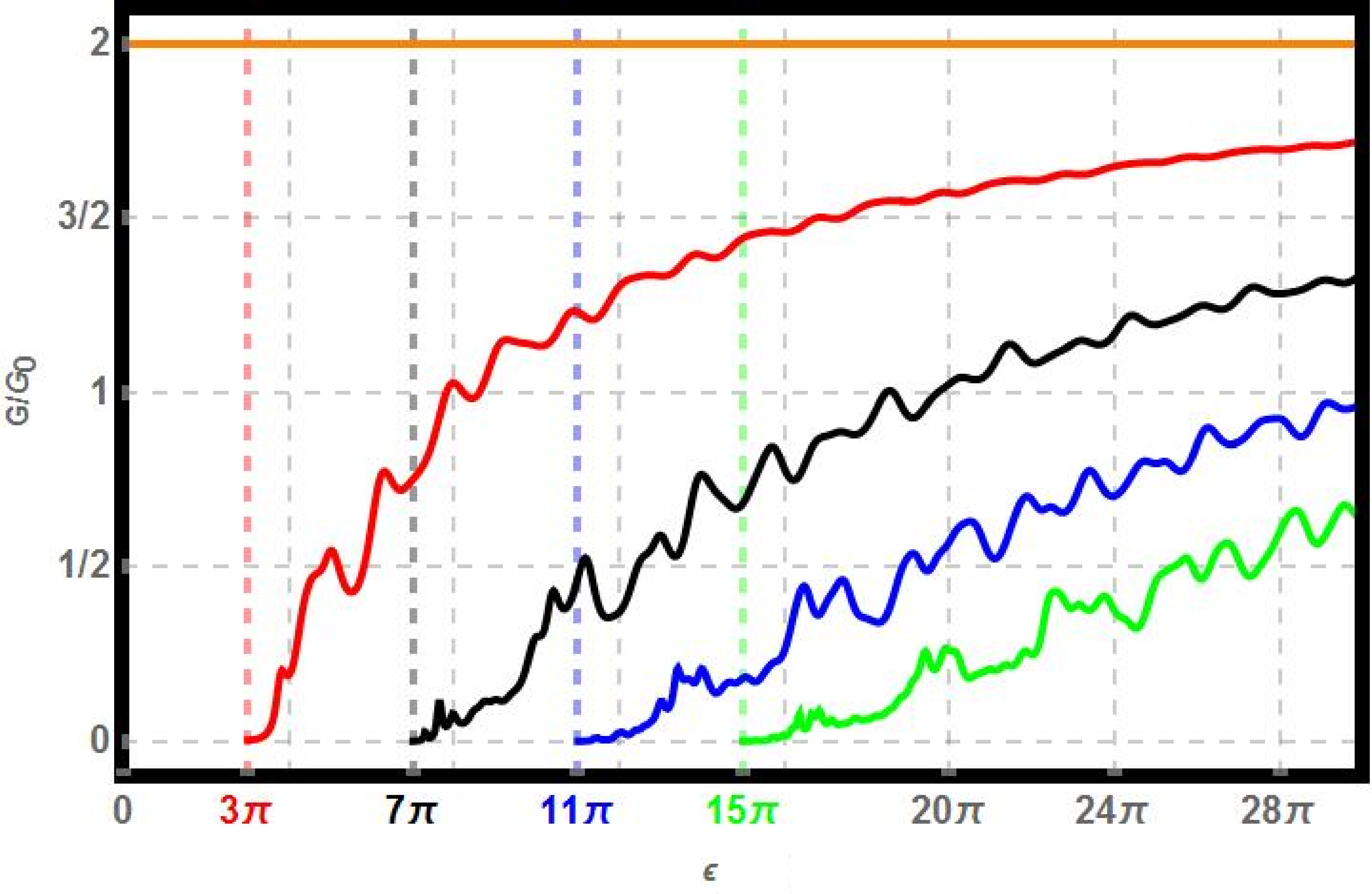}
        \label{fig10:SubFigC}
    }\hspace{-0.02cm}
        \subfloat[]{
        \includegraphics[scale=0.15]{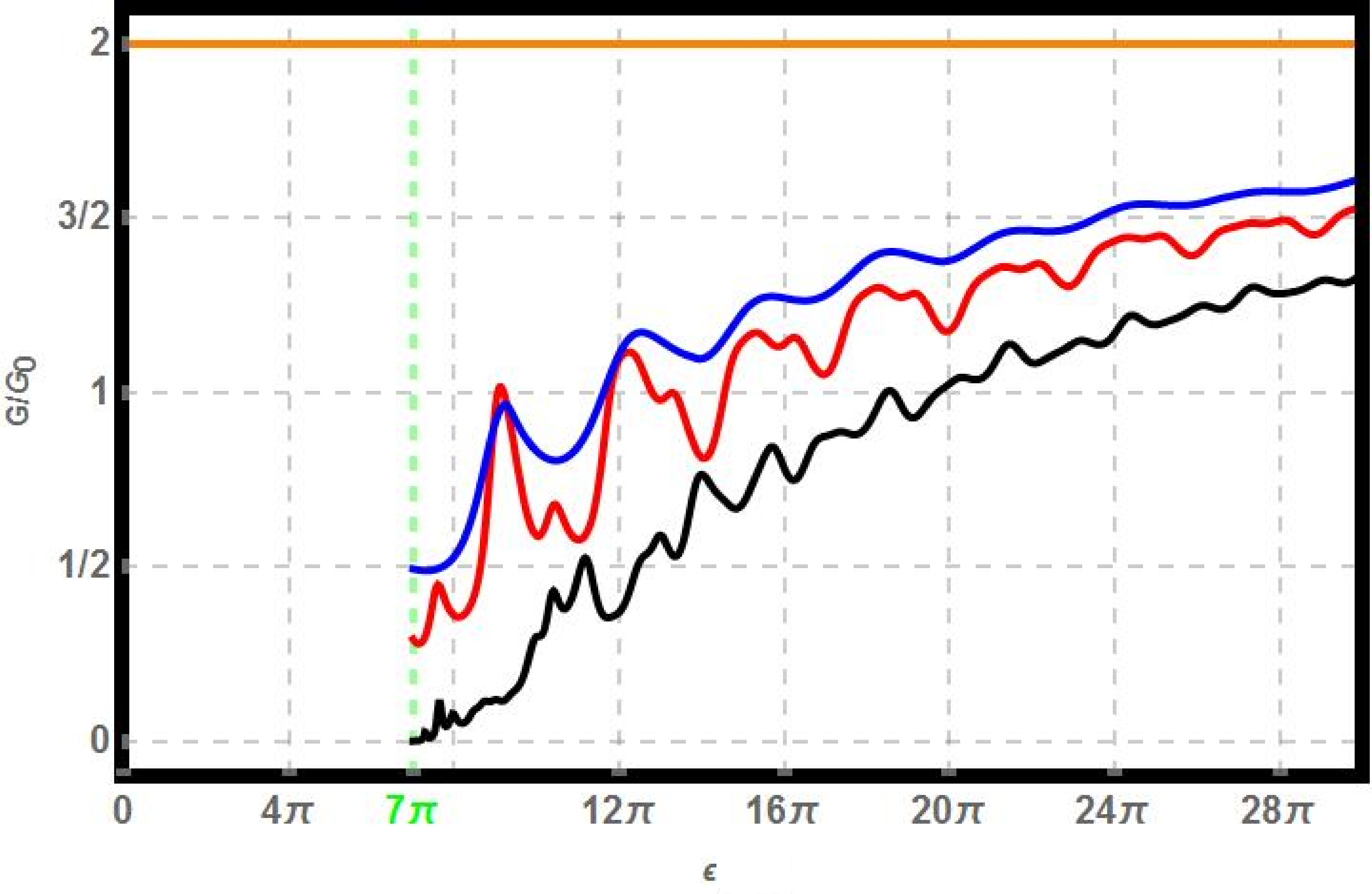}
        \label{fig10:SubFigD}
    }
    \caption{
                (color online) Conductance (in units of $G_0$)  $\frac{G}{G_0}$ versus incident energy
                $\varepsilon$
                for different configurations of
                strained double barrier,
with
                $\varrho=300$, $d=3\varrho$, $\mathbb{V}_2=\mathbb{V}_4=0$.
                \protect \subref{fig10:SubFigA}:
                Strained single barrier with $\delta \tau_2=\pm\delta \tau, \delta \tau_3=\delta \tau_4=0$
                or $\delta \tau_2= \delta \tau_3=0,\delta \tau_4=\pm\delta \tau$ where $\delta \tau=0$
                (orange line), $\delta \tau=3\pi$ (red line), $\delta \tau=7\pi$ (black line), $\delta \tau=11\pi$
                (blue line) and $\delta \tau=15\pi$ (green line).
                \protect \subref{fig10:SubFigB}:
                Strained double barrier with  $\delta \tau_2=\delta \tau_4=\pm\delta \tau, \delta \tau_3=0$
                where $\delta \tau=0$ (orange line), $\delta \tau=3\pi$ (red line), $\delta \tau=7\pi$
                (black line), $\delta \tau=11\pi$ (blue line) and $\delta \tau=15\pi$ (green line).
\protect \subref{fig10:SubFigC}:
                Strained double barrier with  $\delta \tau_2=-\delta \tau_4=\pm\delta \tau, \delta \tau_3=0$
                where $\delta \tau=0$ (orange line), $\delta \tau=3\pi$ (red line), $\delta \tau=7\pi$ (black line),
                $\delta \tau=11\pi$ (blue line) and $\delta \tau=15\pi$ (green line).
\protect \subref{fig10:SubFigD}:
                Strained double barrier for $\delta \tau=7\pi$ in the cases:
                ($\delta \tau_2= \delta \tau_4=\pm\delta \tau,\delta \tau_3=0$) (red line),
                ($\delta \tau_2= -\delta \tau_4=\pm\delta \tau,\delta \tau_3=0$) (back line),
                (${\delta \tau_2=\pm\delta \tau, \delta \tau_3= \delta\tau_4=0}$) or
                ($\delta \tau_2=\delta \tau_3=0, {\delta \tau_4}=\pm\delta \tau$) (blue line).
        }
        \label{Fig10}
\end{figure}

At this level, we show that the corresponding conductance is
affected by physical parameters and depends
on the nature of system. Indeed,
Figure \ref{Fig10} shows the conductance as a function of incident energy $\varepsilon$ for
different configurations of a strained double barrier which is not subject to any potential, i.e.
($\mathbb{V}_2=\mathbb{V}_4=0$).
In Figures (\ref{Fig10}\subref{fig10:SubFigA}, \ref{Fig10}\subref{fig10:SubFigB},
 \ref{Fig10}\subref{fig10:SubFigC}), we present the conductance for strained single barrier
(($\delta \tau_2=\pm\delta \tau, \delta \tau_3=\delta \tau_4=0$) or
($\delta \tau_2= \delta \tau_3=0,\delta \tau_4=\pm\delta \tau$)),
double barrier with the same strain ($\delta \tau_2=\delta \tau_4=\pm\delta \tau, \delta \tau_3=0$)
and double barrier with different strains ($\delta \tau_2=-\delta \tau_4=\pm\delta \tau, \delta \tau_3=0$),
respectively. We observe that under some  physical conditions
there are conductances for $\delta \tau=0$
(orange color), $\delta \tau=3\pi$ (red color), $\delta \tau=7\pi$ (black color),
$\delta \tau=11\pi$ (blue color) and $\delta \tau=15\pi$ (green color). These  Figures
show that each conductance starts at the energy $\varepsilon= \delta \tau$ and there is an ordering such that
$G/G_0(\delta \tau=0)>G/G_0(\delta \tau=3\pi)>G/G_0(\delta \tau=7\pi)>G/G_0(\delta \tau=11\pi)>G/G_0(\delta \tau=15\pi)$.
Figure \ref{Fig10}\subref{fig10:SubFigD} illustrates the comparison between conductances corresponding to
strained single barrier (SSB) (blue color) and
strained double barriers for both:
same strain (SSDB) (red color) and  different strains (DSDB) (black color)
with $\delta \tau=7\pi$. Then it is clearly seen that 
the result $G/G_0$ (SSB) $>$ $G/G_0$ (SSDB) $>$ $G/G_0$ (DSDB) holds.

\begin{figure}[!ht]\centering
    \subfloat[]{
        \includegraphics[scale=0.15]{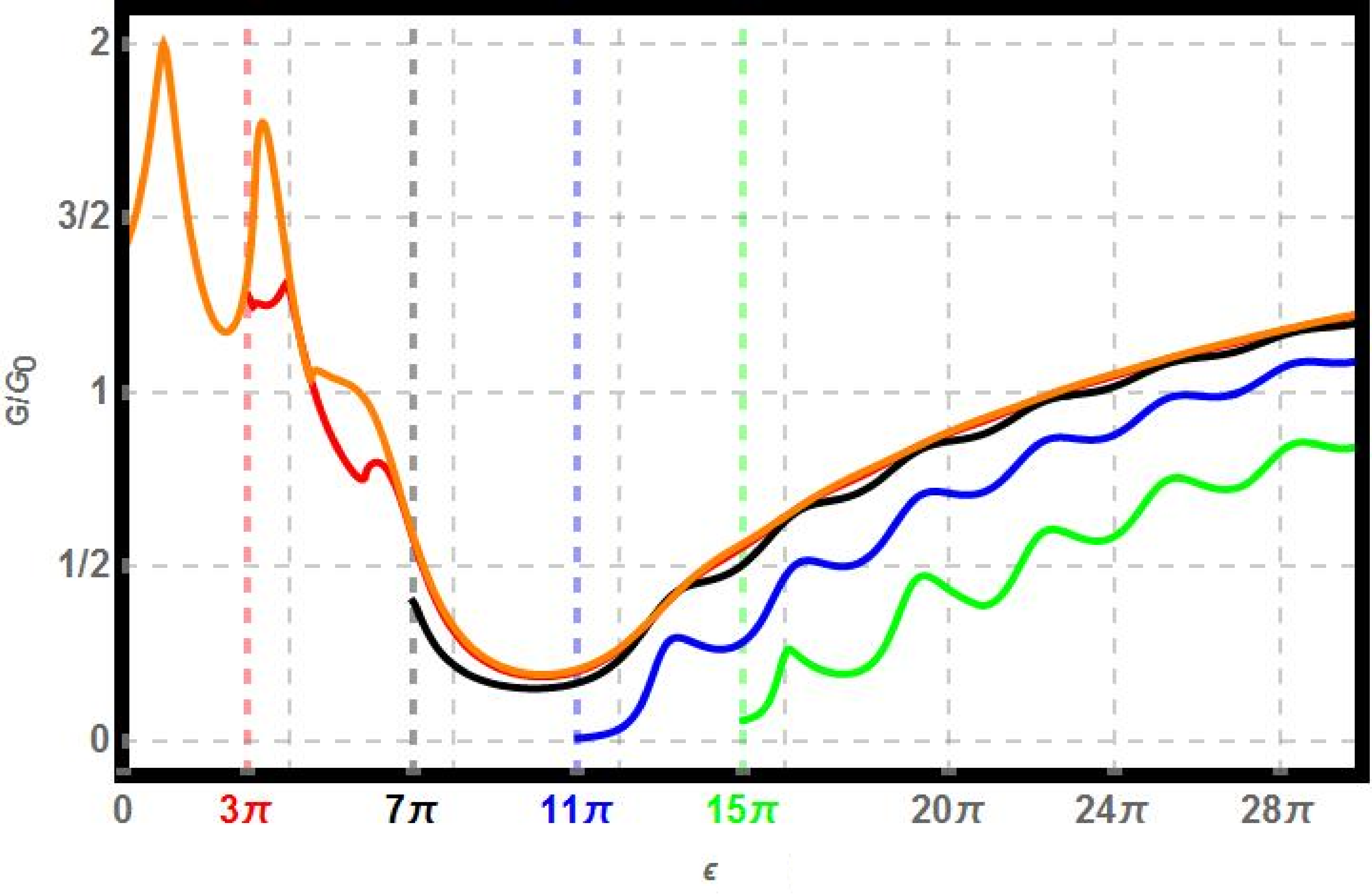}
        \label{fig11:SubFigA}
    }\hspace{-0.02cm}
        \subfloat[]{
        \includegraphics[scale=0.15]{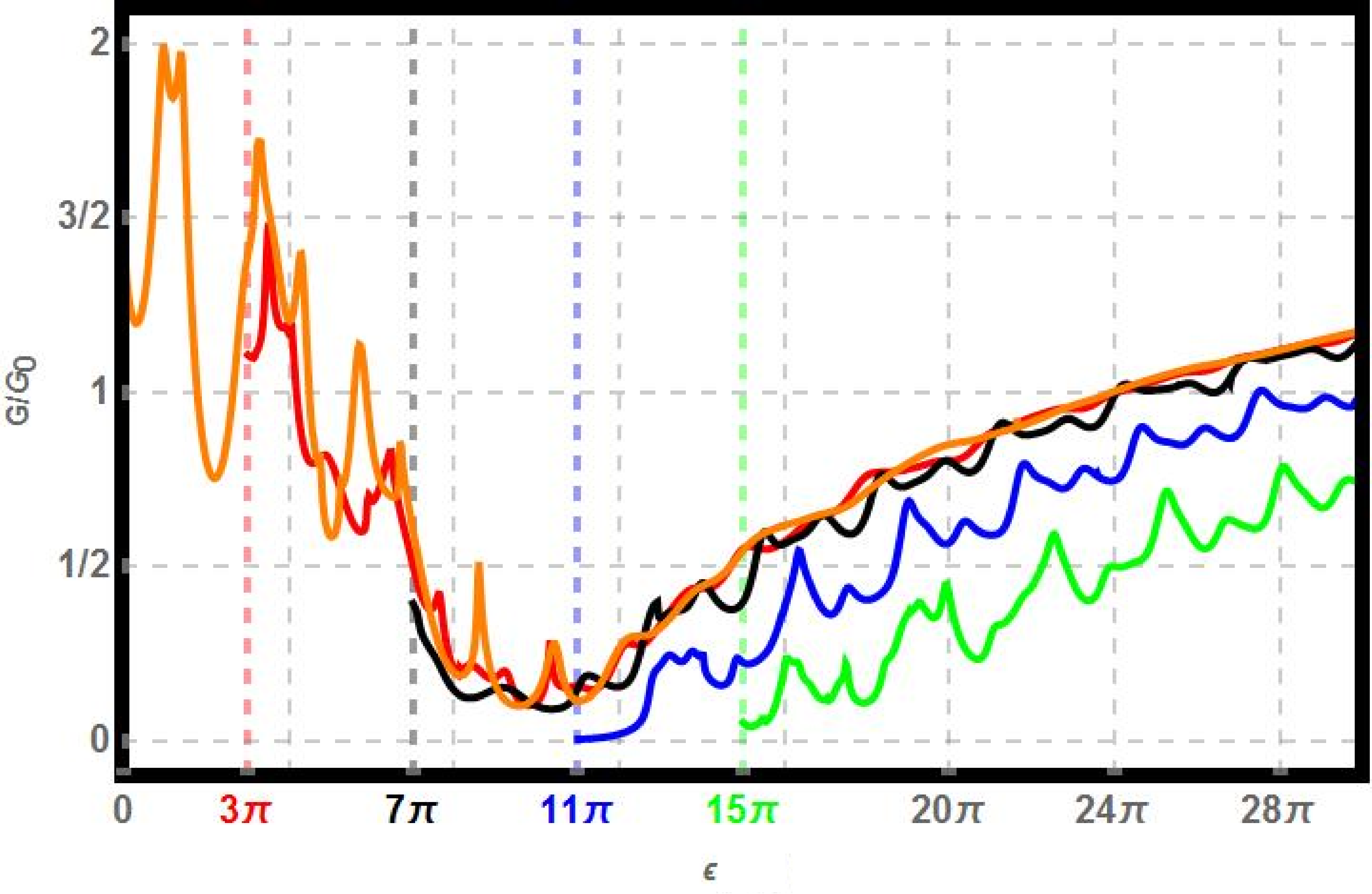}
        \label{fig11:SubFigB}
    }\\
        \subfloat[]{
        \includegraphics[scale=0.15]{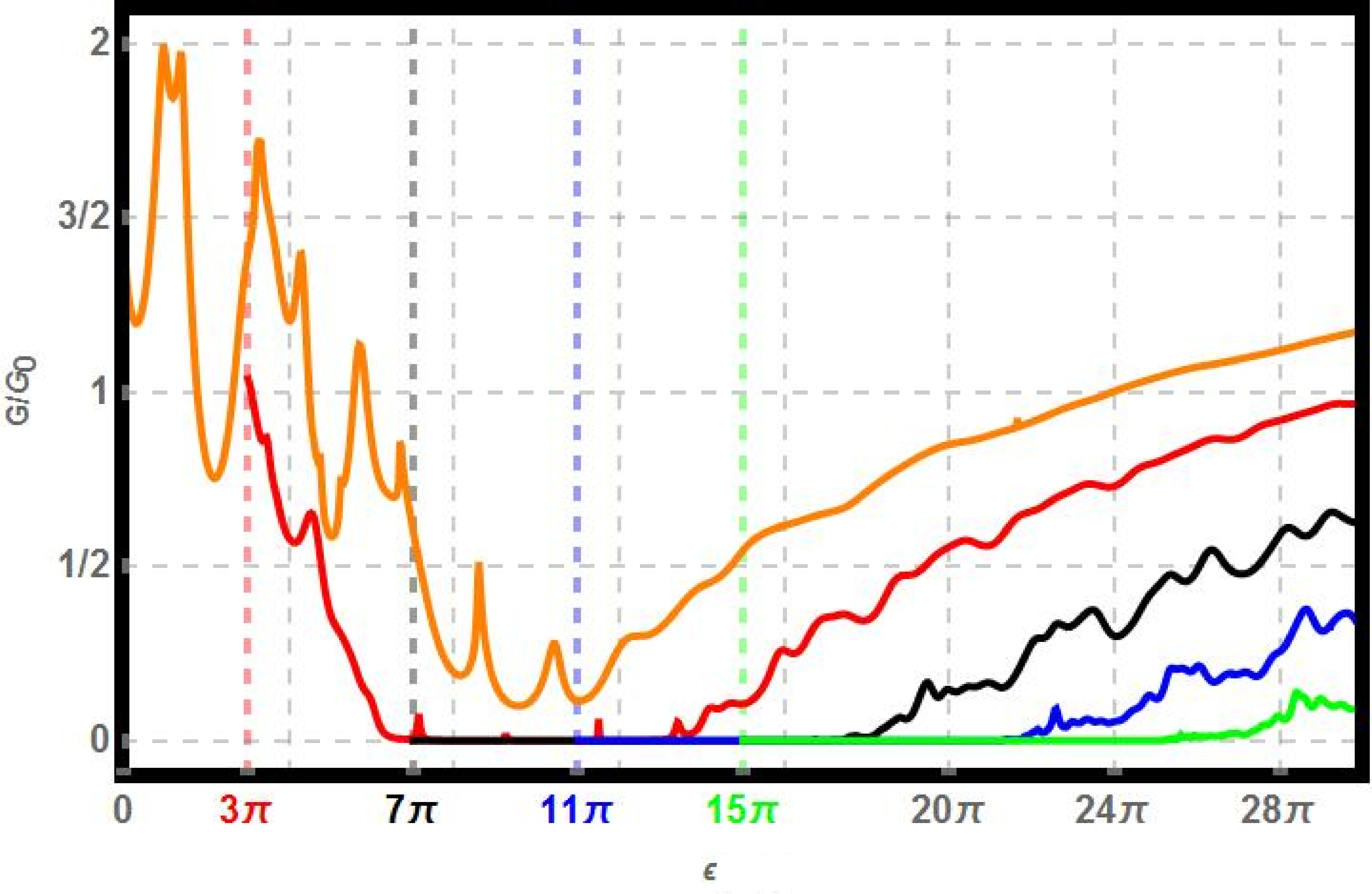}
        \label{fig11:SubFigC}
    }\hspace{-0.02cm}
        \subfloat[]{
        \includegraphics[scale=0.15]{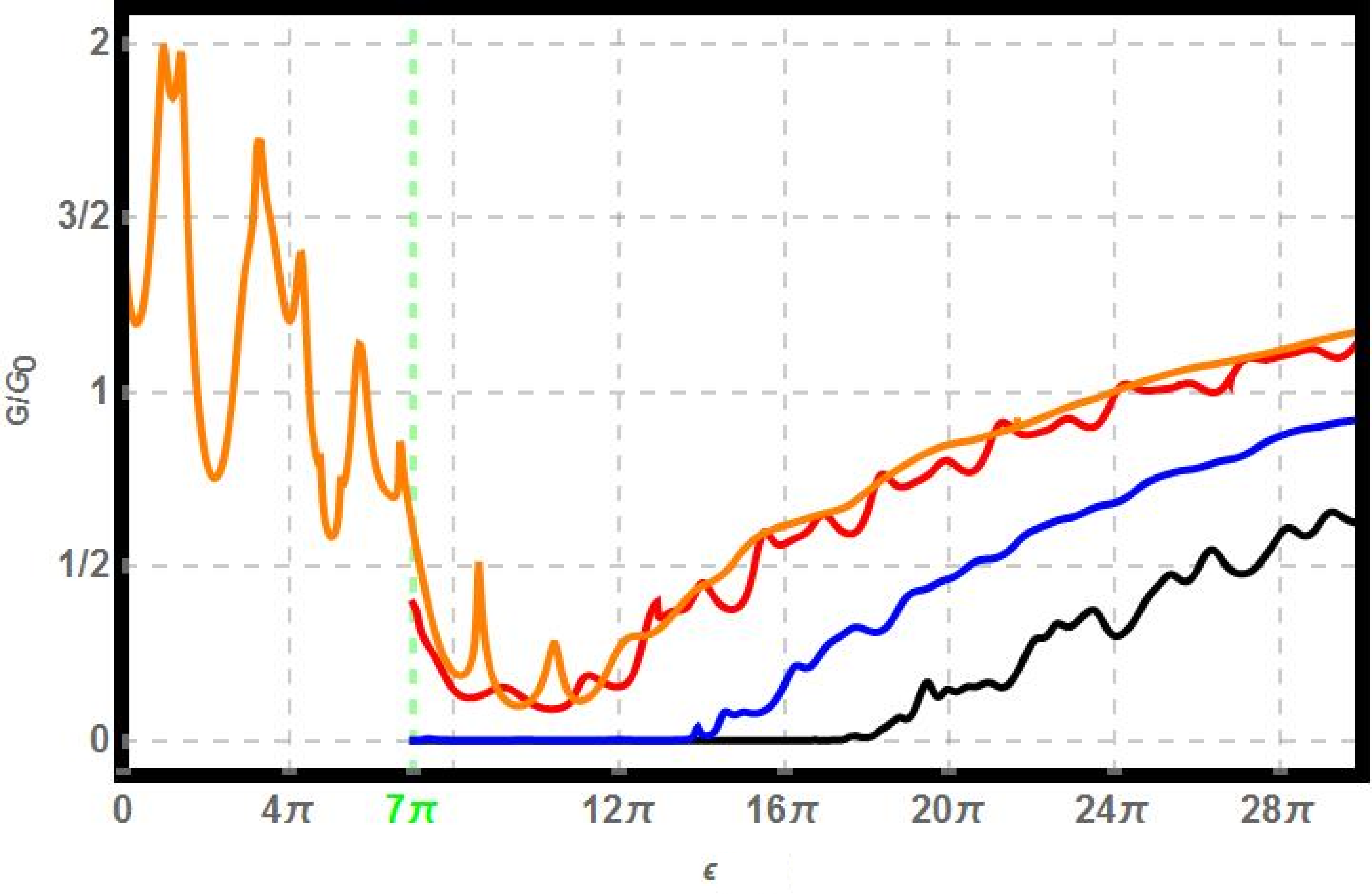}
        \label{fig11:SubFigD}
    }
    \caption{(color online) Conductance (in units of $G_0$)
    $\frac{G}{G_0}$ versus incident energy
    for different configurations of strained double barrier,
                $\varepsilon$ 
with
                $\varrho=300$, $d=3\varrho$ and ($\mathbb{V}_2=\mathbb{V}_4=10\pi$).
                \protect \subref{fig11:SubFigA}:
                Strained single barrier, ($\delta \tau_2=\pm\delta \tau, \delta \tau_3=\delta \tau_4=0$) or
                ($\delta \tau_2= \delta \tau_3=0,\delta \tau_4=\pm\delta \tau$) where $\delta \tau=0$ (orange line),
                $\delta \tau=3\pi$ (red line), $\delta \tau=7\pi$ (black line), $\delta \tau=11\pi$ (blue line) and
                $\delta \tau=15\pi$ (green line).
                \protect \subref{fig11:SubFigB}:
                Strained double barrier with  ($\delta \tau_2=\delta \tau_4=\pm\delta \tau, \delta \tau_3=0$)
                where $\delta \tau=0$ (orange line), $\delta \tau=3\pi$ (red line), $\delta \tau=7\pi$
                (black line), $\delta \tau=11\pi$ (blue line) and $\delta \tau=15\pi$ (green line).
\protect \subref{fig11:SubFigC}:
                Strained double barrier with  ($\delta \tau_2=-\delta \tau_4=\pm\delta \tau, \delta \tau_3=0$)
                where $\delta \tau=0$ (orange line), $\delta \tau=3\pi$ (red line), $\delta \tau=7\pi$ (black line),
                $\delta \tau=11\pi$ (blue line) and $\delta \tau=15\pi$ (green line).
\protect \subref{fig11:SubFigD}:
                Strained double barrier for $\delta \tau=7\pi$ in case:
                ($\delta \tau_2= \delta \tau_4=\pm\delta \tau,\delta \tau_3=0$) (red line),
                ($\delta \tau_2= -\delta \tau_4=\pm\delta \tau,\delta \tau_3=0$) (back line),
                (${\delta \tau_2=\pm\delta \tau, \delta \tau_3= \delta\tau_4=0}$) or
                ($\delta \tau_2=\delta \tau_3=0, {\delta \tau_4}=\pm\delta \tau$) (blue line).
        }
        \label{Fig11}
\end{figure}

Figures \ref{Fig11} shows the effect of strain together with
applied potential on the conductances
with ($\mathbb{V}_2=\mathbb{V}_4=10\pi$) and ($\mathbb{V}_2=10\pi,
\mathbb{V}_4=0$ or $\mathbb{V}_2=0,\mathbb{V}_4=10\pi$). 
{We observe that the behavior of the conductance  changed absolutely compared to Figure \ref{Fig10}.
The strainless double barrier conductance in 
Figures \ref{Fig11}\subref{fig11:SubFigB}, \ref{Fig11}\subref{fig11:SubFigC},
\ref{Fig11}\subref{fig11:SubFigD} and 
single barrier conductance in Figure
\ref{Fig11}\subref{fig11:SubFigA} (orange color) have even minimum located at
$\varepsilon=\mathbb{V}=10\pi$. 
These two conductances differ in resonance peaks, which due to the double barrier (Fabry-Perot
resonator model). Figures \ref{Fig11}\subref{fig11:SubFigA} (SSB) and
\ref{Fig11}\subref{fig11:SubFigB} (SSDB) show that the conductances, for
$\delta \tau=0$ (orange color), $\delta \tau=3\pi$ (red color), $\delta \tau=7\pi$ (black color),
are almost comparable in values and  only differ in their mass terms corresponding to the limit of
the forbidden zones, respectively,  $\varepsilon=0,3\pi$ and $7\pi$.
But the conductances for $\delta \tau=11\pi$ (blue color) and $\delta \tau=15\pi$ (green color)
remain separated and only differs by oscillations}.
In Figure \ref{Fig11}\subref{fig11:SubFigC}  of DSDB the three conductances fell at
 energy $\varepsilon=\mathbb{V}=\delta \tau$  until the value zero,
then they resume after an increase in energy.
A comparison between the three kinds of strained double barriers subject to a
potential $\mathbb{V}$ is illustrated in Figure \ref{Fig11}\subref{fig11:SubFigC}.
The conductance of  SSB has the same minimum as that of simple potential barrier
$\mathbb{V}=10\pi$ at $\varepsilon=\mathbb{V}=10\pi$, but the conductance of the SSDB and
DSDB start with zero at $\varepsilon=\mathbb{V}=\delta \tau$. By increasing their Fermi energies,
SSB resumes its conductance before DSDB.

\section{Conclusion}

We have studied the electronic structure of Dirac fermions through a double barrier potential
in a strained graphene ribbon. Our system  is a graphene
chip made up of five regions where the second and fourth  subjected to
the strain effect and double barrier potential.
In such system the armchair configuration  along the $x$-direction imposes  a gauge field $\bm{\mathcal{A}}_j$
perpendicular to the direction of  propagation, which only affected horizontal inter-carbon links.
After writing the Hamiltonian governing Dirac fermions in each region, we have determined
the energies spectrum and associated eigenspinors.  It was
shown that the transverse wave
vector is shifted by the strain term compared to that of the pristine graphene and
the strain directly affects the transmission angle of Dirac fermions.
The continuity of the eigenspinors at each interface separating
two consecutive regions, allowed us to build a  transfer matrix connecting the  amplitudes of propagation
in the \emph{input} and \emph{output} regions.
We have calculated the
transmission and reflection probabilities together with  the conductance at zero temperature.

Subsequently, we have numerically presented 
the transmission probability density plots
versus the incident energy $\varepsilon$ and incident angle $\theta$ under suitable conditions.
These density  plots  showed  the forbidden and permitted zones of transmission delimited by
the energy contours ${\varepsilon=\frac{\delta \tau}{1\pm\sin\theta}}$, where
Klein paradox zones are highlighted in red  (total transmission), transmission gaps in purple
and  forbidden zones in white in all generated Figures.  It has been seen  that the Klein paradox  is not always verified
at normal  incident angle and the transmissions satisfy three symmetry relations related to
the incident angle and strain effect (deformation), see  
(\ref{24a}-\ref{24c}). 
 The
Fermi surfaces, corresponding to the dispersion relation,  allowed us to determine the collimation angles
for different configurations of the strain effect and double barrier.
These angles can also be obtained using the conservation  of the $k_y$ wave-vector component throughout
the Dirac fermions propagation. We have showed 
that the transmission probability exhibited double resonance peaks, which
can be explained by Fabry-Perot resonator model.
At  grazing incidence angle $\theta=\frac{\pi}{2}$,  we have seen that the transmission showed only resonance peaks and  transmission gaps.

By considering 
double barrier potential together with strain effect, we have showed
that
the transmission behavior
changed completely.
Indeed, the allowed and forbidden zones  of transmission were now
delimited by energy contours 
{$\varepsilon^{s}_{s'}= \frac{\mathbb{V}+s\delta\tau}{ 1+ s'\sin\theta}$} with $s=\pm$ and $s'=\mp$.
Always under the conservation constraint   of the $k_y$ wave-vector component,
and taking into account the Fermi surfaces,  we have the following correspondences
between the delimiting energies contours and the collimation angles:
{$\varepsilon^{s}_{s'}= \frac{\mathbb{V}+\alpha\delta\tau}{ 1+ s'\sin\theta}\rightarrow \theta^{s}_{s'}=\arcsin\left( \frac{\mathbb{V}-\varepsilon+s\delta\tau}{s'\varepsilon}\right) $}
Finally, we have performed a
comparative study of the conductances between different configurations of
strained double barrier with or without potential profile.

\section*{Acknowledgments}

The generous support provided by the Saudi Center for Theoretical
Physics (SCTP) is highly appreciated by all authors. AJ and HB
acknowledge the support of KFUPM under research group project RG181001.


\begin{thebibliography}{99}



\bibitem{6} J. Scott Bunch, Arend M. van der Zande, Scott S. Verbridge, Ian W. Frank, David M. Tanenbaum,
Jeevak M. Parpia, Harold G. Craighead, Paul L. McEuen, Science 315, 490 (2007).
\bibitem{7} N. Levy, S. A. Burke, K. L. Meaker, M. Panlasigui, A. Zettl, F. Guinea, A. H. Castro Neto,
M. F. Crommie, Science 329, 544 (2010).

 \bibitem{8} V. M. Pereira and A. H. Castro Neto, Phys. Rev. Lett. 103,
046801 (2009).
\bibitem{9} F. Guinea, M. I. Katsnelson and A. K. Geim, Nat. Phys. 6, 30
(2010).



\bibitem{10} F. De Juan, A. Cortijo, M. A. H. Vozmediano, A. Cano, Nat. Phys. 7, 810 (2011).

\bibitem{3} K. S. Novoselov, A. K. Geim, S. V. Morozov, D. Jiang, Y. Zhang, S. V. Dubonos,
I. V. Grigorieva and A. A. Firsov, Science 306, 666 (2004).
\bibitem{4} K. S. Novoselov, A. K. Geim, S. V. Morozov, D. Jiang, M. I. Katsnelson, I. V. Grigorieva,
S. V. Dubonos and A. A. Firsov, Nature 438, 197 (2005).
\bibitem{5} Yuanbo Zhang, Yan-Wen Tan, Horst L. St\"ormer and Philip Kim, Nature 438, 201 (2005).
\bibitem{Castro}
A. H. Castro Neto, F. Guinea, N. M. R. Peres, K. S. Novoselov, and A. K. Geim, Rev. Mod. Phys. 81, 109 (2009)

\bibitem{Bahlouli}
H. Bahlouli, E. B. Choubabi, A. Jellal, and M. Mekkaoui,   J. Low Temp. Phys. 169,  51 (2012).

\bibitem{AIP} C. Yesilyurt, S. Ghee Tan, G. Liang, and  M. B. Jalil,  AIP Advances 6 (5), 056303 (2016).

\bibitem{Buttiker1985} M. B\"uttiker, Y. Imry, R. Landauer, and S. Pinhas, Phys. Rev. B 31, 6207 (1985).
\bibitem{zah1} X. Xia, J. Wang, F. Zhang, Z. D. Hu, C. Liu, X. Yan, and L. Yuan, Plasmonics, 10, 6, 1409(2015).
\bibitem{zah2} N. Gu, M. Rudner, and  L. Levitov, Phys. Rev. lett. 107, 156603 (2011).
\bibitem{zah3}  A. Daboussi, L. Mandhour, J. N. Fuchs, and  S. Jaziri,  Phys. Rev. B 89,  085426 (2014).
  \bibitem{Chinese} S. Li-Feng, D. Li-Min, W. Zhi-Fang, and  F. Chao, Chinese Phys. B 22,  077201 (2013).
  \bibitem{Peres}    N. M. R. Peres, A. C. Neto, and F. Guinea,   Phys. Rev. B 73, 195411 (2006).



\end{thebibliography}
\end{document}